\documentclass[aps,prl,twocolumn,superscriptaddress,floatfix]{revtex4-2}
\usepackage{graphicx}
\usepackage{multirow}
\usepackage{amsmath}
\usepackage{amsfonts}
\usepackage{amssymb}
\graphicspath{{Figures/}}
\usepackage{epstopdf}
\usepackage{empheq}
\usepackage{float}
\usepackage{cases}
\usepackage{mathtools}
\usepackage{dcolumn}
\usepackage{bbold}
\usepackage{xfrac}
\usepackage{bm}
\usepackage{siunitx}
\usepackage{dsfont}
\usepackage{physics}
\usepackage[dvipsnames]{xcolor}
\usepackage[colorlinks]{hyperref}
\usepackage[normalem]{ulem}
\usepackage{orcidlink}
\usepackage{gensymb}

\hypersetup{
colorlinks=True,
linkbordercolor=White,
linkcolor=BrickRed,
citecolor=Blue
}

\renewcommand{\vec}[1]{\boldsymbol{#1}}

\newcommand{\PRLsep}{\noindent\makebox[\linewidth]{\resizebox{0.5\linewidth}{1pt}{$\bullet$}}\bigskip}
\newcommand{\decorate}[1]{\bar{#1}}

\begin{document}

\title{Spin-cQED with bulk germanium spin qubits}

\author{A.-F. Kalo}
\affiliation{Univ. Grenoble Alpes, CEA, IRIG-MEM-L\_Sim, Grenoble, France}
\author{E. A. Rodríguez-Mena}
\affiliation{Univ. Grenoble Alpes, CEA, IRIG-MEM-L\_Sim, Grenoble, France}
\author{J. C. Abadillo-Uriel$^{\orcidlink{0000-0001-6411-2544}}$}
\affiliation{Quantum Advanced Research Center (QuARC) - Consejo Superior de Investigaciones Cient{\'i}ficas (CSIC)}
\affiliation{Instituto de Ciencia de Materiales de Madrid (ICMM) - Consejo Superior de Investigaciones Cient{\'i}ficas (CSIC), Sor Juana In{\'e}s de la Cruz 3, 28049 Madrid, Spain}
\author{M. Filippone$^{\orcidlink{0000-0003-1102-8335}}$}
\affiliation{Univ. Grenoble Alpes, CEA, IRIG-MEM-L\_Sim, Grenoble, France}
\author{Y.-M. Niquet$^{\orcidlink{0000-0002-1846-1396}}$}
\affiliation{Univ. Grenoble Alpes, CEA, IRIG-MEM-L\_Sim, Grenoble, France}

\begin{abstract}
Unstrained bulk germanium is a particularly attractive material for circuit quantum electrodynamics with spins (spin-cQED). We show, through systematic modeling and comparison with state-of-the-art strained germanium heterostructures, that hole spins in bulk germanium double quantum dots readily reach the strong-coupling regime with superconducting microwave resonators, achieving spin-photon coupling strengths $g_s/2\pi\gtrsim100$\,MHz. This enhancement originates from large spin-orbit interactions beyond the perturbative regime. In addition, the coupling is much less sensitive to the orientation of the applied magnetic field, which shall ease operation and limit the impact of device-to-device variability. Our results establish bulk germanium as a compelling platform for scalable spin-cQED.
\end{abstract}

\maketitle
\maketitle{\it Introduction --} Spin circuit quantum electrodynamics (spin-cQED) studies the coherent interaction between microwave photons and the spins of charge carriers in solid-state devices~\cite{Clerk2020}. Building on cavity QED~\cite{haroche2006exploring} then superconducting circuit QED~\cite{Blais2021}, spin-cQED replaces superconducting artificial atoms with spins, the most elementary coherent two-level systems~\cite{burkard2023semiconductor}.

However, microwave photons strongly couple to the charge~\cite{mi2017strong,Stockklauser2017,bruhat_2018_cQED,scarlino2021situ,granel20263dintegrationhybridquantum,van2026dispersivereadoutsimosquantum}, but not directly to the spin of the carriers. Engineering sufficient spin-charge hybridization for coherent spin-photon interfaces therefore remains a major experimental and theoretical challenge~\cite{Imamoglu1999,Viennot2015,Samkharadze2018,Mi2018,Landig2018,borjans2020resonant,harvey-collard2022coherent}. Nevertheless, the strong intrinsic spin-orbit (SO) interactions of holes in silicon have recently enabled spin-photon coupling strengths exceeding $g_s/2\pi=100$\,MHz~\cite{yu2022strong} while preserving spin coherence~\cite{noirot2026coherence}. Beyond their fundamental interest, such interfaces can mediate coherent interactions between distant spins and provide a route to fast dispersive spin readout~\cite{d2019optimal,borjans2020resonant,harvey-collard2022coherent,kam2024submicrosecond,dijkema2025cavity}.

Strained germanium heterostructures are now the state-of-the-art material for SO-based electrical hole spin manipulation~\cite{Scappucci20,Hendrickx20b,Hendrickx20,Hendrickx21,Valentin25}. Spin-photon interfaces on this platform are, however, far less advanced~\cite{de2024strong,kang2024coupling,janik2025strong,depalma2025lowlossfrequencytunablejosephsonjunction,ruggiero2026high}. Whether strained germanium can achieve figures of merit comparable to silicon remains an open question. The main challenges in this material are the small gate lever arms~\cite{Hendrickx20b,Borsoi24,Hendrickx2024,Valentin25}, and the strong anisotropy of the gyromagnetic tensor and weaker SO interactions due to lattice-mismatch strain~\cite{Terrazos21,Wang21,Michal21,Bosco21b,martinez2022hole,Abadillo2023,Mauro25,sarkar2025effect}.

\begin{figure}
    \includegraphics[width=1\columnwidth]{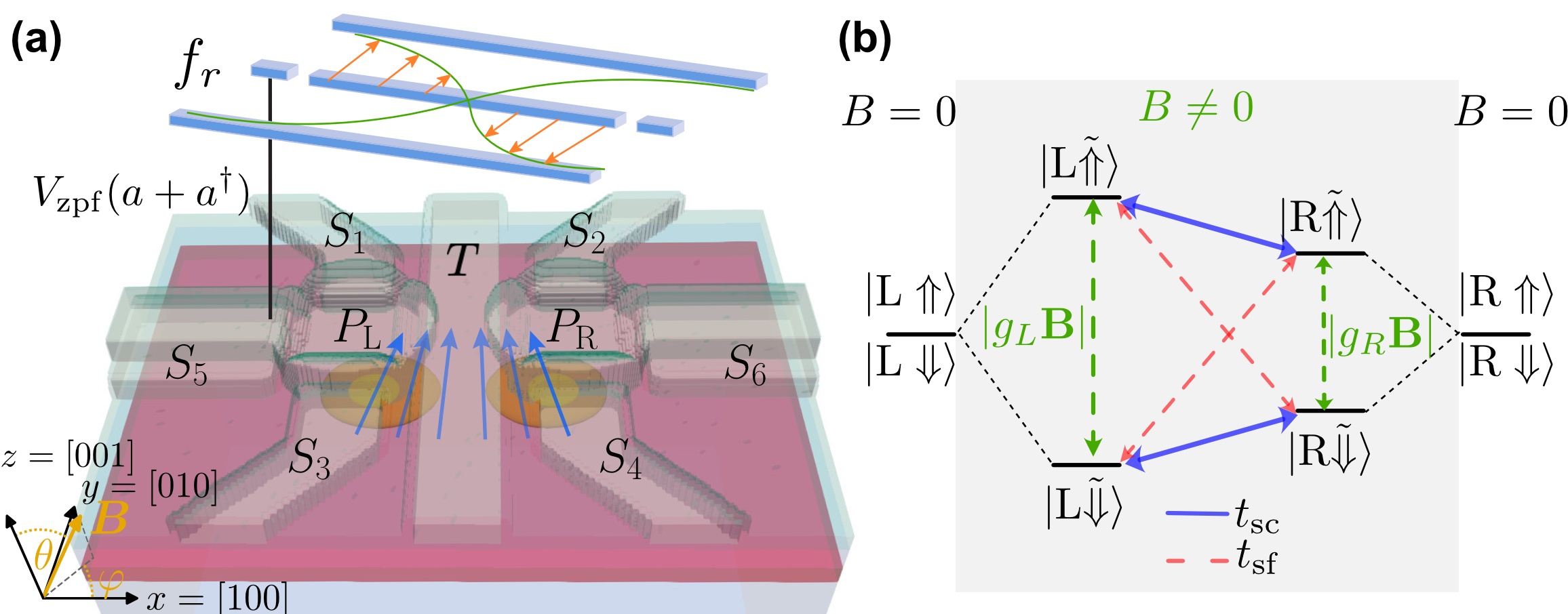}
    \caption{Device for spin-photon coupling. (a) The strained QW device. Germanium is colored in red, Ge$_{0.8}$Si$_{0.2}$ in blue and the aluminium gates in gray. The 20-nm-thick gates are arranged on two levels separated from each other and from the heterostructure by 6\,nm of Al$_2$O$_3$: the barrier ($T$) and side gates ($S_n$) on the first level, and the plunger gates $P_\mathrm{L}$ and $P_\mathrm{R}$ on the second (diameter 130\,nm; distance between centers 170\,nm). A resonator with frequency $f_r$ and zero-point fluctuations $V_\mathrm{zpf}$ is connected to $P_\mathrm{L}$. In the bulk heterostructure the buffer and QW are replaced by an unstrained Ge substrate. The yellow shapes are the iso-density surfaces that enclose 20\% and 80\% of the L and R orbitals. The orientation of the magnetic field $\vec{B}$ is characterized by the angles $\theta$ and $\varphi$. (b) Under the magnetic field $\vec{B}$, the degenerate spins $\Uparrow/\Downarrow$ of the L/R orbitals split into $\ket*{{\rm L/R}\tilde{\Uparrow}}$ and $\ket*{{\rm L/R}\tilde{\Downarrow}}$ states according to their respective $g$-matrices $g_\mathrm{L/R}$. These states are then hybridized by spin-conserving ($t_\mathrm{sc}$) and SO mediated spin-flip tunneling ($t_\mathrm{sf}$) processes~\cite{yu2022strong}. This spin-charge hybridization enables spin-photon coupling and is pictured by the blue arrows in (a).}
    \label{fig:device}
\end{figure}

Unstrained bulk germanium has recently emerged as a compelling alternative that may overcome most of these limitations~\cite{Scappucci25,Mauro2025b}. Here, we systematically model and compare spin-photon coupling in double-quantum-dot (DQD) devices hosted in strained and bulk germanium (Fig.~\ref{fig:device}). We show that holes in bulk germanium DQDs can readily achieve much larger spin-photon couplings ($g_s/2\pi\gtrsim100$\,MHz) owing to strong SO interactions beyond the standard perturbative regime. We further show that $g_s$ is much less sensitive to the orientation of the applied magnetic field, which shall ease operation and scalability. These results establish bulk germanium as a competitive platform for solid-state spin-cQED.

{\it Setup and model --} We consider the prototypical device of Fig.~\ref{fig:device}(a). The ``strained'' heterostructure comprises a 16-nm-thick Ge quantum well (QW) biaxially strained by a Ge$_{0.8}$Si$_{0.2}$ buffer and capped with a 20-nm-thick Ge$_{0.8}$Si$_{0.2}$ barrier. The biaxial strains in the well are, therefore, $\varepsilon_{xx}=\varepsilon_{yy}=-0.61\%$ and $\varepsilon_{zz}=0.45\%$ \cite{Mauro2025b}. The ``unstrained'' bulk heterostructure is a Ge substrate capped with the same Ge$_{0.8}$Si$_{0.2}$ barrier~\cite{Mauro2025b}. The DQD is shaped by the difference of potential between two plunger gates $P_\mathrm{L}$ and $P_\mathrm{R}$, the barrier gate $T$ and the side gates $S_n$. Both heterostructures actually undergo additional (but small) inhomogeneous strains imprinted by the thermal contraction of these gates upon cool-down \cite{Abadillo2023}.

We model these devices using the numerical framework of Refs.~\cite{martinez2022hole,Abadillo2023} and supplemental material (SM)~\cite{SM}. The motion of the hole is described by the Luttinger–Kohn (LK) Hamiltonian \cite{Luttinger56,Winkler03} that mixes heavy-hole (HH) and light-hole (LH) states. This mixing controls the SO interactions that enable spin-photon coupling, and results from the interplay between the confinement in the potential of the gates, the magnetic field $\vec{B}$, and the inhomogeneous cool-down strains.

The physics of the device can be captured by a minimal Hamiltonian in a $\{\ket{\rm L\Uparrow},\ket{\rm L\Downarrow},\ket{\rm R\Uparrow},\ket{\rm R\Downarrow}\}$ basis set for the left (L) and right (R) dot spin orbitals. Assuming holes with positive dispersion, $H=H_\mathrm{DQD}+H_\mathrm{r}$, with $H_\mathrm{DQD}=H_\mathrm{S}+H_\mathrm{T}+H_\mathrm{MT}$ the DQD Hamiltonian with Zeeman, tunneling, and magnetotunneling terms \cite{rodriguezmena2025}, and $H_\mathrm{r}$ the resonator Hamiltonian:
\begin{align}
    H_\mathrm{S}&=\frac{\varepsilon}{2}\tau_z+\frac{\mu_B}{2}\left[(\vec{\sigma}\cdot g_\mathrm{L}\vec{B})\tau_\mathrm{L}+(\vec{\sigma}\cdot g_\mathrm{R}\vec{B})\tau_\mathrm{R}\right] \nonumber \\
    H_\mathrm{T}&=-(t_c\cos\theta_\mathrm{so})\tau_x-(t_c\sin\theta_\mathrm{so}\vec{n}_\mathrm{so}\cdot\vec{\sigma})\tau_y \nonumber \\
    H_\mathrm{MT}&=\frac{\mu_B}{2}(\boldsymbol{\sigma}\cdot g_\mathrm{T}\vec{B})\tau_x+\frac{1}{2}(\boldsymbol{\mu}_\mathrm{T}\cdot\vec{B})\tau_y \nonumber \\
    H_\mathrm{r}&=\hbar\omega_ra^\dagger a+eV_{\rm zpf}(a+a^\dagger)D_\mathrm{L}\,.
    \label{eq:H}
\end{align}
Here $\vec{\sigma}$ is the vector of Pauli matrices acting in the $\{\ket{\Uparrow},\ket{\Downarrow}\}$ subspace, and $\tau_\alpha$ are the Pauli matrices acting in the $\{\ket{\rm L},\ket{\rm R}\}$ subspace, with $\tau_\mathrm{L,R}=\tfrac{1}{2}(\tau_0\pm\tau_z)$ and $\tau_0$ the identity. The uncoupled L and R dots are characterized by their detuning energy $\varepsilon$ and by their gyromagnetic $g$-matrices $g_\mathrm{L}$ and $g_\mathrm{R}$~\cite{Venitucci18}. The spin thus precesses around the axis $\vec{\omega}_\mathrm{L,R}=\mu_B g_\mathrm{L,R}\vec{B}/\hbar$ (at angular frequency $|\vec{\omega}_\mathrm{L,R}|$) in the Bloch spheres of the L and R dots (with $\mu_B$ the Bohr magneton). The tunneling Hamiltonian $H_\mathrm{T}$ features a spin-independent ($\propto\tau_x$) term and a SO contribution (which rotates the spin of a tunneling hole by an angle $2\theta_\mathrm{so}$ about vector $\vec{n}_\mathrm{so}$). The magnetotunneling Hamiltonian $H_\mathrm{MT}$ collects all linear-in-$\mathbf{B}$ corrections to $H_\mathrm{T}$ through the matrix $g_\mathrm{T}$ and vector $\boldsymbol{\mu}_T$~\cite{rodriguezmena2025}. The last line of Eq.~\eqref{eq:H} couples the DQD to a resonator with frequency $f_r=\omega_r/2\pi$ connected to gate $P_\mathrm{L}$ ($a$ being the photon annihilation operator). The zero-point fluctuations of this resonator (with amplitude $V_\mathrm{zpf}$) drive the hole through the operator $D_\mathrm{L}\equiv\partial V_t(\vec{r})/\partial V_\mathrm{L}$, the derivative of the total potential in the DQD with respect to the voltage on $P_\mathrm{L}$. In the above minimal Hamiltonian, $D_\mathrm{L}\equiv\alpha_\mathrm{L}\tau_\mathrm{L}$, with $\alpha_\mathrm{L}\approx 0.22$ the lever arm of gate $P_\mathrm{L}$.

\begin{figure*}[!t]
    \centering
    \includegraphics[width=.95\textwidth]{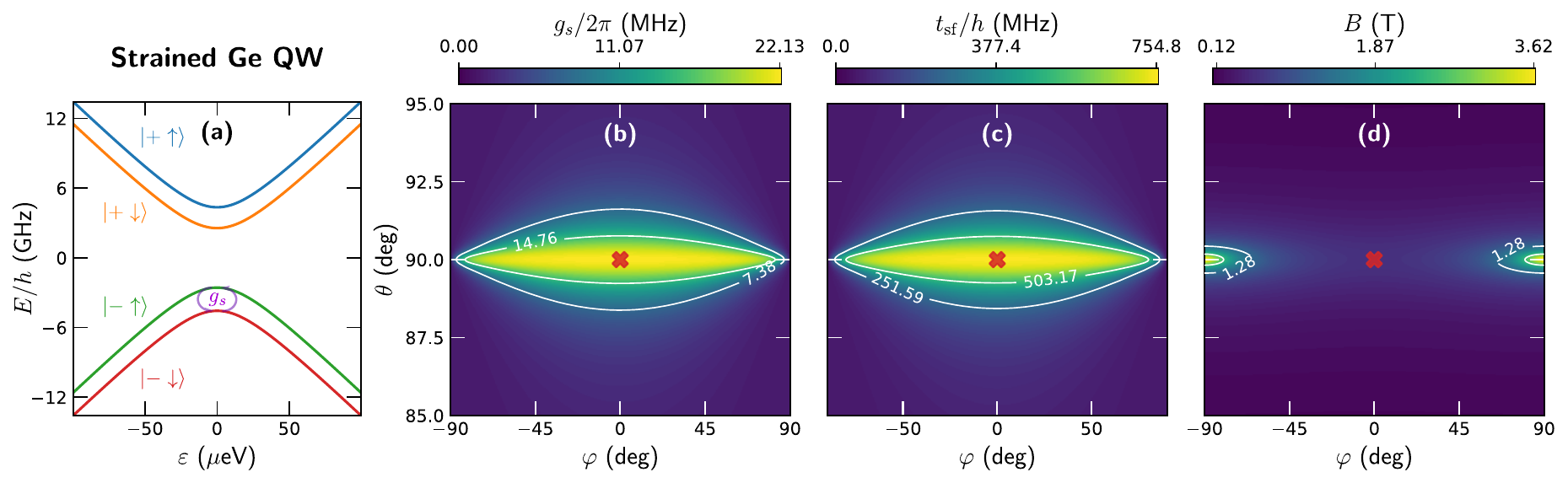}
    \includegraphics[width=.95\textwidth]{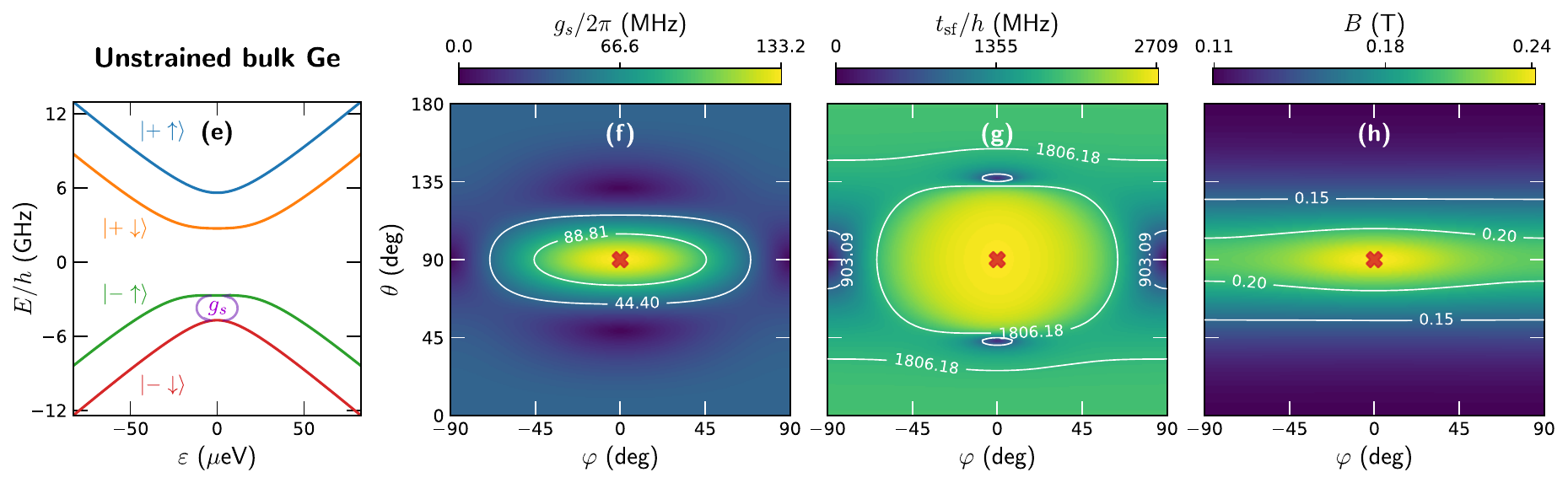}
    \caption{Spin-photon coupling in (a-d) a strained Ge QW and (e-h) an unstrained, bulk Ge heterostructure: (a, e) Spectrum of the DQD as a function of detuning energy $\varepsilon$ ($\vec{B}\parallel\vec{x}$). The resonator couples to the hole in the $\ket{-\uparrow}$ and $\ket{-\downarrow}$ states; (b, f) Spin-photon coupling $g_s$, (c, g) spin-flip tunnel coupling $t_\mathrm{sf}$, (d, h) magnetic field amplitude at resonance ($\omega_s/2\pi=2$\,GHz) as a function of the orientation of $\vec{B}$ (see Fig.~\ref{fig:device}). The red cross highlights the optimal orientation $\vec{B}\parallel\vec{x}$. In the strained Ge QW, $V_\mathrm{L}=V_\mathrm{R}=-20.5$\,mV and $V_\mathrm{T}=-5$\,mV (all side gates grounded), while in the bulk Ge heterostructure, $V_\mathrm{L}=V_\mathrm{R}=-43.4$\,mV and $V_\mathrm{T}=-11.6$\,mV (top and bottom side gates bias $V_{\mathrm{S}_1}=V_{\mathrm{S}_2}=V_{\mathrm{S}_3}=V_{\mathrm{S}_4}=30$\,mV).}
    \label{fig:combined}
\end{figure*}

We emphasize that the parameters $g_\mathrm{L,R,T}$, $\vec{n}_\mathrm{so}$, $\theta_\mathrm{so}$ and $\boldsymbol{\mu}_\mathrm{T}$ depend on the choice of the $\Uparrow,\,\Downarrow$ basis states in each dot. While observables such hole energies or spin-photon couplings do not depend on this choice, the physical interpretation of these parameters does. We fix this gauge by constructing a ``canonical'' heavy-hole-like basis set in the framework of first-order perturbation theory in the HH/LH mixing (see SM~\cite{SM}). The above parameters can then be analyzed in this framework, which enables, e.g., the discussion of effective Rashba and Dresselhaus SO interactions.

At high detuning $|\varepsilon|\gg t_c$, the hole is localized in either the left ($V_\mathrm{L}\ll V_\mathrm{R}$) or right ($V_\mathrm{L}\gg V_\mathrm{R}$) dot. Near zero detuning, the $\ket{\rm L}$ and $\ket{\rm R}$ orbitals are hybridized by the charge tunnel coupling $t_c$, which opens a gap $2t_c$ between the bonding and anti-bonding states $\ket{\pm}$. They are further split by the magnetic field into $\ket{\pm\uparrow}$ and $\ket{\pm\downarrow}$ states, as illustrated in Figs.~\ref{fig:combined}a,e. In the presence of SO coupling, $\uparrow$ and $\downarrow$ are generally mixtures of $\Uparrow$ and $\Downarrow$, so that the resonator can drive transitions between the ground $\ket{-\uparrow}$ and $\ket{-\downarrow}$ states. This results in an effective spin-photon Hamiltonian for this doublet~\cite{cottet2015electron-photon,michal2022tunable,Bosco22,Sagaseta_2026}:
\begin{align}
    H_\text{eff}&=\hbar\omega_r a^\dagger a+\frac{\hbar\omega_s}{2}\sigma_z+\hbar g_\parallel(a+a^\dagger)\sigma_z+\hbar g_s(a+a^\dagger)\sigma_x \nonumber \\ \hbar g_\parallel &=eV_\mathrm{zpf}\left[\bra{-\uparrow}D_{\rm L}\ket{-\uparrow}-\bra{-\downarrow}D_{\rm L}\ket{-\downarrow}\right]/2 \nonumber \\ 
    \hbar g_s&=eV_\mathrm{zpf}|\bra{-\uparrow}D_{\rm L}\ket{-\downarrow}|
    \,,
    \label{eq:gs}
\end{align}
where $\omega_s=(E_{-\uparrow}-E_{-\downarrow})/\hbar$ is the spin angular frequency, $g_\parallel$ is the longitudinal, and $g_s$ the transverse spin-photon coupling. In the following we operate at the sweet spot $\varepsilon=0$ of symmetric DQDs, where $\omega_s$ is first-order insensitive to detuning; $g_\parallel$ thus vanishes and we focus on the transverse coupling $g_s$.

To understand the effects of SO coupling, it is insightful to switch to a local-Zeeman frame~\cite{yu2022strong}, as illustrated in Fig.~\ref{fig:device}b. Leaving aside $H_\mathrm{MT}$, we can diagonalize the Zeeman Hamiltonians $H_Z^\mathrm{L/R}=\frac{\mu_B}{2}\vec{\sigma}\cdot g_\mathrm{L/R}\vec{B}$ of both dots, and introduce the spin-flip tunnel matrix element $t_\mathrm{sf}=\bra*{\rm L\tilde{\Uparrow}}H_\mathrm{T}\ket*{\rm R\tilde{\Downarrow}}=-\bra*{\rm L\tilde{\Downarrow}}H_\mathrm{T}\ket*{\rm R\tilde{\Uparrow}}$~\footnote{In contrast to the spin-flip term, there is a spin-conserving tunnel coupling $t_\mathrm{sc}=-\bra*{\rm L\tilde{\Uparrow}}H_\mathrm{T}\ket*{\rm R\tilde{\Uparrow}}=-\bra*{\rm L\tilde{\Downarrow}}H_\mathrm{T}\ket*{\rm R\tilde{\Downarrow}}$, such that $t_c^2=t_\mathrm{sc}^2+t_\mathrm{sf}^2$.} between the Zeeman-split $\{\ket*{\rm L\tilde{\Uparrow}},\ket*{\rm L\tilde{\Downarrow}}\}$ and $\{\ket*{\rm R\tilde{\Uparrow}},\ket*{\rm R\tilde{\Downarrow}}\}$ states (which depend on the orientation of the magnetic field). At low magnetic field amplitude ($\hbar\omega_s\ll t_c$), the spin-photon coupling
\begin{equation}
    \hbar g_s\approx \frac{1}{4}e\alpha_\mathrm{L} V_\mathrm{zpf}\frac{\hbar(\omega_\mathrm{L}+\omega_\mathrm{R})}{2t_c}\frac{t_\mathrm{sf}}{t_c}
    \label{eq:gspert}
\end{equation}
is actually proportional to $t_\mathrm{sf}$. In the absence of SO tunneling terms ($\theta_\mathrm{so}=0$), it can be shown that~\cite{Geyer24}
\begin{equation}
    t_\mathrm{sf}=t_c\sin(\Theta_\mathrm{LR}/2)\le t_c
    \label{eq:tsf}
\end{equation}
with $\Theta_\mathrm{LR}$ the angle between the Larmor vectors $\vec{\omega}_\mathrm{L}\propto g_\mathrm{L}\vec{B}$ and $\vec{\omega}_\mathrm{R}\propto g_\mathrm{R}\vec{B}$. If $\theta_\mathrm{so}\ne 0$, Eq.~\eqref{eq:tsf} still holds with the angle $\widehat{\Theta}_\mathrm{LR}$ between the vectors $\hat{\vec{\omega}}_\mathrm{L}={\cal R}(-\theta_\mathrm{so},\vec{n}_\mathrm{so})\vec{\omega}_\mathrm{L}$ and $\hat{\vec{\omega}}_\mathrm{R}={\cal R}(+\theta_\mathrm{so},\vec{n}_\mathrm{so})\vec{\omega}_\mathrm{R}$, where ${\cal R}(\theta,\vec{u})$ is the rotation of angle $\theta$ about $\vec{u}$. Spin-flip tunneling thus arises not only from a finite $\theta_\mathrm{so}$ but also from the imbalance between the $g$-matrices of the two dots (both contributions ultimately resulting from SO coupling in the system). For more general expressions including magnetotunneling corrections, see the SM~\cite{SM}.

For all heterostructures, we fit the parameters of the effective Hamiltonian \eqref{eq:H} on the eigensolutions of a finite-difference implementation of the LK Hamiltonian. We tune $V_\mathrm{L/R/T}$ so that the charge tunnel coupling is $t_c/h=3.5$\,GHz. We assume that the resonator frequency is $f_r=2$\,GHz and that $V_\mathrm{zpf}=10\,\mu$V~\cite{yu2022strong}. We compute the spin-photon coupling with Eq.~\eqref{eq:gs} and the eigenstates of Eq.~\eqref{eq:H} (without the resonator). We adjust the magnetic field amplitude to achieve the spin/photon resonance condition $\omega_s=2\pi f_r$.

{\it Strained QW device --} We start with a symmetric DQD in a strained heterostructure. The spectrum of the DQD (Fig.~\ref{fig:combined}a) highlights the expected tunneling gap between the L and R orbitals (shown in Fig.~\ref{fig:device}). The spin-photon coupling $g_s/2\pi$ at zero detuning and the spin-flip tunnel matrix element $t_\mathrm{sf}$ are plotted as a function of the orientation of $\vec{B}$ in Fig.~\ref{fig:combined}b,c. $g_s/2\pi$ peaks when $\vec{B}\parallel\vec{x}$ but reaches at best 22\,MHz. The tunneling Hamiltonian $H_\mathrm{T}$ is almost diagonal. The small $\theta_\mathrm{so}=0.83\degree$ about $\vec{n}_\mathrm{so}=\vec{y}$ is the signature of the (primarily cubic) Rashba interaction emerging in such quasi-circular quantum dots \cite{Marcellina17,Terrazos21,Rodriguez2023}, however strongly suppressed by the large HH/LH bandgap $\Delta\approx 70$\,meV opened by the biaxial strains~\cite{Mauro25}. Therefore, $g_s$ essentially results from the mismatch between the $g$-matrices of the dots. Their diagonals are actually the same ($g_{xx}=0.234$, $g_{yy}=-0.046$, $g_{zz}=13.78$) and they mostly differ by opposite $g_{zx}=\pm 0.048$. The strong anisotropy of the diagonal elements (reflected in magnetic field strength needed to reach $\omega_s/2\pi=2$\,GHz, Fig.~\ref{fig:combined}d) is characteristic of almost pure HH states (LH mixing $<0.15\%$). The off-diagonal $g_{zx}$ describes a rotation of the principal gyromagnetic $x$ and $z$ axes by an angle $\delta\theta\approx g_{zx}/g_{zz}$ around $\vec{y}$~\cite{Mauro25}. This tilt results from the coupling between the in- and out-of-plane motions of the hole in the non-separable potential of the DQD~\cite{martinez2022hole}, and from the shear strains $\varepsilon_{xz}$ imposed by the thermal contraction of the metal gates~\cite{Abadillo2023}. When the magnetic field is in-plane, the precession axis of the spin thus rocks out-of-plane when the hole tunnels from one dot to the other, which gives rise to spin-flip transitions. Nevertheless, $t_\mathrm{sf}\lesssim 0.22t_c$ is far from optimal. Moreover, the in-plane peak of $g_s$ is very thin [full width at half maximum (FWHM) $\Delta\theta\approx 2.25\degree$] owing to the large gyromagnetic anisotropy. The precession axes of the spin in the L and R dots indeed lock onto $z$ once the magnetic field goes slightly out-of-plane, which suppresses dot-to-dot modulations. This calls for a careful alignment of the magnetic field and may complicate long-range coupling between distant DQDs that may be misoriented with respect to each other.

The spin-photon coupling can be strengthened in asymmetric DQDs thanks to the stronger imbalance between the $g$-matrices of the dots~\cite{MRodriguez2025}. $g_s$ then still peaks (almost) in-plane, at an azimuthal angle $\varphi\ne 0$ that takes best advantage of the mismatch between the in-plane $g$ factors $g_{xx}$ and $g_{yy}$, but strongly depends on the geometry of the dots. Nevertheless, $g_s/2\pi$ hardly exceeds $50$\,MHz at reasonable magnetic fields $B<1$\,T in our simulations, far below the best values measured in fully-depleted silicon-on-insulator devices ($g_s/2\pi=330$\,MHz at $f_r=5.43$\,GHz and similar $\hbar\omega_s/t_c\approx 0.57$ \cite{yu2022strong}). We emphasize, though, that the lever arm is, by design, smaller in germanium than in silicon devices with thin SiO$_2$ gate oxides ($\alpha_\mathrm{L}\approx 0.5$ in \cite{yu2022strong}). Therefore, the zero-point voltage fluctuations of the resonator are more strongly coupled to the DQD in silicon.

\begin{figure}[!t]
    \centering
    \includegraphics[width=.95\columnwidth]{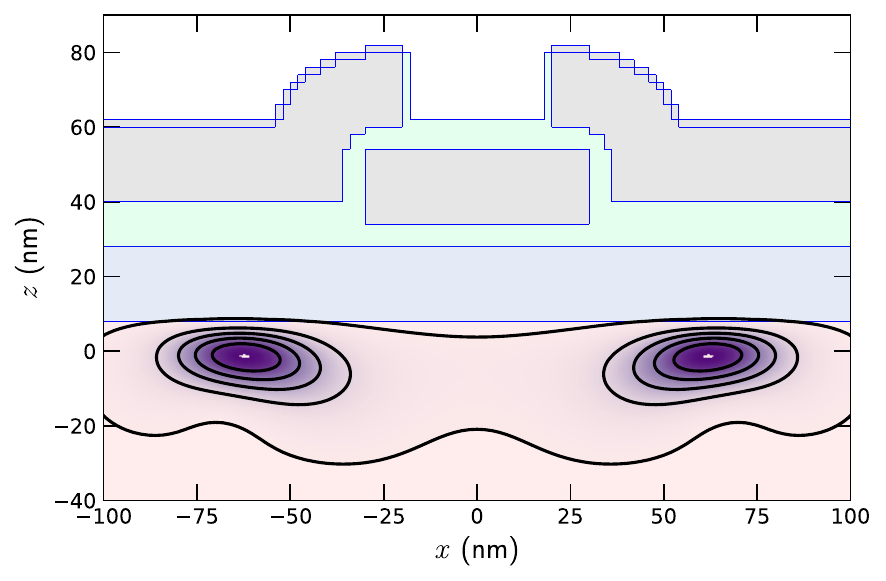}
    \caption{Map of the density (arbitrary units) in the $\ket{-}$ state of the unstrained, bulk Ge heterostructure at zero detuning, in the $(xz)$ symmetry plane of the device.}
    \label{fig:wfnsunstrained}
\end{figure}

{\it Unstrained bulk device --} We can compare these data with those calculated in unstrained bulk Ge, plotted in Fig.~\ref{fig:combined}e-h. The dots are still symmetric but are $35\%$ larger along $x$ than along $y$ due to the bias on the side gates. Similar maps can nevertheless be obtained for quasi-circular and even asymmetric dots (see SM~\cite{SM}). The maximum $g_s/2\pi\approx 133$\,MHz when $\vec{B}\parallel\vec{x}$ is about 6$\times$ larger than in the strained QW. The in-plane features are also much more robust and broader (FWHM $\Delta\theta\approx 39\degree$), which enables easier optimization of the operation point and shall limit the impact of variability. This results from a weaker gyromagnetic anisotropy ($g_{xx}=0.454$, $g_{yy}=0.694$, $g_{zz}=1.086$), promoted by a stronger HH/LH mixing ($\mathrm{LH}=15\%$) enhanced by the smaller HH/LH bandgap $\Delta=1.73$\,meV (now controlled by confinement in the electric field of the gates). The gyromagnetic matrices of the L and R dots exhibit larger, opposite tilts of the principal axes ($g_{zx}=\pm 1.18$ and $g_{xz}=\pm 0.571$). The wave functions of the DQD (Fig.~\ref{fig:wfnsunstrained}) indeed show the fingerprints of strong coupling between the in- and out-of-plane motions due to the weaker confinement along $z$ (with a pronounced rotation of these envelopes in the $(xz)$ plane). Moreover, the tunneling Hamiltonian $H_\mathrm{T}$ is highly non-diagonal ($\theta_\mathrm{so}=60.8\degree$ around $\vec{n}_\mathrm{so}=\vec{y}$), which suggests the emergence of strong Rashba SO interactions with characteristic length scales comparable to the inter-dot distance. The Rashba coefficients calculated in the one- and two-dimensional limits are actually up to three orders of magnitude larger in bulk Ge than in strained QWs (see SM~\cite{SM})~\cite{Bosco21b,delvecchio2026}. We emphasize, though, that generic Rashba interactions have limited relevance in bulk Ge due to the large HH/LH mixing (beyond the perturbative regime) and strong coupling between the in- and out-of-plane motions. Magnetotunneling (another fingerprint of strong SO coupling) also has a significant impact on bulk devices (here, $H_\mathrm{MT}$ enhances $g_s$ by $\approx 20\%$) \cite{rodriguezmena2025}. Overall, SO coupling is so strong that $g_s(\vec{B}\parallel\vec{x})$ would be slightly larger if $\theta_\mathrm{so}$ were zero because the net tunneling spin rotation $\widehat{\Theta}_\mathrm{LR}=\Theta_\mathrm{LR}+2\theta_\mathrm{so}$ [resulting from the combined action of the $g$-matrices and $\theta_\mathrm{so}$ in Eq.~\eqref{eq:tsf}] is farther from $180\degree$ than $\Theta_\mathrm{LR}$ ($\Theta_\mathrm{LR}=137.91\degree$ and $\widehat{\Theta}_\mathrm{LR}=259.45\degree$). Nevertheless, $t_\mathrm{sf}$ reaches $0.77t_c$ and remains, therefore, almost optimal.

{\it Conclusions --} In this work, we have carried out a comparative study of strained and bulk Ge quantum dot devices for spin-cQED. The spin-photon coupling per effective zero-point voltage fluctuations, $g_s/(2\pi\alpha_\mathrm{L}V_\mathrm{zpf})\approx 60$\,MHz/$\mu$V, is much larger in bulk than in strained Ge, and is comparable to the best values measured in silicon ($\approx 74$\,MHz/$\mu$V) \cite{yu2022strong}, enabling the strong  coupling regime. The spin-flip tunnel matrix elements calculated in bulk Ge are, indeed, quasi-optimal ($t_\mathrm{sf}\gtrsim 0.75t_c$), so that $g_s$ is primarily limited by the lever arm of the gate the resonator is connected to. Moreover, there are several ways to optimize these devices and potentially reach the ultrastrong coupling regime $g_s\gtrsim 0.1\omega_r$. First, the lever arm may be increased by gate stack engineering, using, e.g., tip-shaped gates embedded in the GeSi barrier \cite{Martinez24}. Second, the SO coupling can be slightly reduced if needed by burying a thin (5\,nm thick) layer of GeSi $\approx 20$\,nm below the surface of the heterostructure (to structurally confine the wave function, see SM~\cite{SM}). Finally, high-quality superconducting resonators, difficult to manufacture on Ge, can be designed on sapphire and bonded to the Ge device using a ``flip-chip'' technology~\cite{granel20263dintegrationhybridquantum}. Unstrained, bulk Ge heterostructures therefore appear as a compelling platform for spin-cQED, with applications such as photon-mediated long-range spin-spin interactions~\cite{dijkema2025cavity} and quantum non-demolition readout~\cite{Blais2021}.

This work was supported by the French National Research Agency (project InGeQT), and by the Horizon Europe Framework Program (grant agreement 101174557 QLSI2). JCAU also acknowledges the support of the Spanish Ministry of Science, Innovation, and Universities through Grants RYC2022-037527-I and PID2023-148257NA-I00 funded by MICIU/AEI/10.13039/501100011033 and by FSE+ and FEDER, UE, ``ERDF A way of making Europe'' and European Union Next Generation EU/PRTR.

\setcounter{section}{0}
\setcounter{equation}{0}
\setcounter{figure}{0}
\setcounter{table}{0}

\setcounter{secnumdepth}{3}

\renewcommand\thefigure{S\arabic{figure}} 
\renewcommand\theequation{S\arabic{equation}}
\renewcommand{\thetable}{S\arabic{table}}

\onecolumngrid

\vspace{1cm}
\PRLsep

\clearpage

\begin{center}
\textbf{\large Supplementary material for ``Spin-cQED with bulk germanium spin qubits''}
\end{center}

In this supplementary material, we detail in section~\ref{sec:methodology} the methodology used to compute the parameters of the effective Hamiltonian [Eq.~(1) of the main text], then discuss in section~\ref{sec:parameters} these parameters (and the underlying physics) for the devices of the main text (as well as additional configurations). We then assess Rashba spin-orbit interactions in bulk germanium in section~\ref{sec:SOC}, and illustrate the possibilities offered by heterostructure engineering in section~\ref{sec:insertion} (with the insertion of a thin GeSi layer below the dots to control spin-orbit coupling). We finally provide general expressions for spin-photon coupling in different frames in section~\ref{sec:frames}.

\section{Methodology}
\label{sec:methodology}

In this section, we present the methodology used to compute the spin-photon coupling in the double quantum dot (DQD). We first introduce the Luttinger-Kohn Hamiltonian, then the effective 4-level Hamiltonian used to rationalize the results, and discuss the definition of the left and right orbitals and pseudo-spins consistent with perturbation theories for the analysis of the physics. We finally describe the Newton-Raphson algorithm used to tune the device to a target tunnel coupling $t_c$.

\subsection{The Luttinger-Kohn Hamiltonian}

In bulk Ge, the highest valence band Bloch functions at $\Gamma$ form a quadruplet that can be mapped onto the $J=\tfrac{3}{2}$ eigenstates of the total angular momentum $\vec{J}=\vec{L}+\vec{S}$. In the minimal $\vec{k}\cdot\vec{p}$ theory \cite{Luttinger55,Luttinger56,Winkler03,KP09}, the wave functions of the hole in the potential $V_t(\vec{r})$ of the gates are expanded as
\begin{equation}
\psi(\vec{r},\sigma_z)=\sum_{\nu\in\left\{-\tfrac{3}{2},-\tfrac{1}{2},\tfrac{1}{2},\tfrac{3}{2}\right\}}\varphi_\nu(\vec{r})u_\nu(\vec{r},\sigma_z)\,,
\label{eq:envelopes}
\end{equation}
where $\sigma_z=\pm 1/2$ is the physical spin, $u_\nu$ is the Bloch function with angular momentum $j_z=\nu$ along $z=[001]$ [$\nu=\pm 3/2$ for the heavy-hole (HH) and $\nu=\pm 1/2$ for the light-hole (LH) components], and $\varphi_\nu$ is an envelope function.

The HH and LH envelope functions $\varphi_\nu$ are the solutions of a set of differential equations defined by the Luttinger-Kohn (LK) Hamiltonian \cite{Luttinger56}
\begin{equation}
H_\mathrm{LK}=H_\mathrm{K}+H_\varepsilon+H_\mathrm{Z}+eV_t(\vec{r})\mathbb{1}_4\,,
\label{eq:H4KP}
\end{equation}
where $H_\mathrm{K}$ is the kinetic energy, $H_\varepsilon$ describes the effects of strains, $H_\mathrm{Z}$ is the Zeeman Hamiltonian and $\mathbb{1}_4$ is the $4\times4$ identity matrix. $H_\mathrm{K}$ and $H_\varepsilon$ share the same generic form in the $j_z=\{+\tfrac{3}{2},+\tfrac{1}{2},-\tfrac{1}{2},-\tfrac{3}{2}\}$ basis set (we consider here holes with positive dispersion):
\begin{equation}
H_{\mathrm{K}/\varepsilon}=\begin{pmatrix}
P+Q & -S & R & 0 \\
-S^\dagger & P-Q & 0 & R \\
R^\dagger & 0 & P-Q & S \\
0 & R^\dagger & S^\dagger & P+Q
\end{pmatrix}
\label{eq:LK}\,,
\end{equation}
where, for $H_\mathrm{K}$,
\begin{subequations}
\label{eq:PQRSkin}
\begin{align}
P_\mathrm{K}&=\frac{1}{2m_0}\gamma_1(p_x^2+p_y^2+p_z^2) \\
Q_\mathrm{K}&=\frac{1}{2m_0}\gamma_2(p_x^2+p_y^2-2p_z^2) \\
R_\mathrm{K}&=\frac{1}{2m_0}\sqrt{3}\left[-\gamma_2(p_x^2-p_y^2)+2i\gamma_3\{p_x,\,p_y\}\right] \\
S_\mathrm{K}&=\frac{1}{2m_0}2\sqrt{3}\gamma_3\{p_x-ip_y,\,p_z\}\,,
\end{align}
\end{subequations}
with $\{A,B\}=\tfrac{1}{2}(AB+BA)$, and, for $H_\varepsilon$,
\begin{subequations}
\label{eq:PQRSstrains}
\begin{align}
P_\varepsilon&=-a_v(\varepsilon_{xx}+\varepsilon_{yy}+\varepsilon_{zz}) \\
Q_\varepsilon&=-\frac{1}{2}b_v(\varepsilon_{xx}+\varepsilon_{yy}-2\varepsilon_{zz}) \\
R_\varepsilon&=\frac{\sqrt{3}}{2}b_v(\varepsilon_{xx}-\varepsilon_{yy})-id_v\varepsilon_{xy} \\
S_\varepsilon&=-d_v(\varepsilon_{xz}-i\varepsilon_{yz})\,.
\end{align}
\end{subequations}
Here $\vec{p}$ is the momentum, $m_0$ is the free electron mass, and $\gamma_1$, $\gamma_2$, $\gamma_3$ are the Luttinger parameters that characterize the hole masses. The $\varepsilon_{\alpha\beta}$ are the strains; $a_v$ is the hydrostatic, $b_v$ the uniaxial and $d_v$ the shear deformation potential of the valence band. The form of Eq.~\eqref{eq:LK}, which couples different $j_z$'s through the $R$ and $S$ terms, embodies the action of spin-orbit (SO) coupling in the valence band. The Zeeman Hamiltonian
\begin{equation}
H_\mathrm{Z}=2\mu_B(\kappa\vec{B}\cdot\vec{J}+q\vec{B}\cdot\vec{J}^3)
\label{eq:Zeeman}
\end{equation}
describes the action of the magnetic field on the Bloch functions, with $\vec{J}\equiv(J_x,J_y,J_z)$ the spin $\tfrac{3}{2}$ operator\footnote{The matrices $J_x$, $J_y$, $J_z$ are:
\begin{equation}
    J_x=\frac{1}{2}\begin{pmatrix}
        0 & \sqrt{3} & 0 & 0 \\
        \sqrt{3} & 0 & 2 & 0 \\
        0 & 2 & 0 & \sqrt{3} \\
        0 & 0 & \sqrt{3} & 0
    \end{pmatrix}\,,
    J_y=\frac{i}{2}\begin{pmatrix}
        0 & -\sqrt{3} & 0 & 0 \\
        \sqrt{3} & 0 & -2 & 0 \\
        0 & 2 & 0 & -\sqrt{3} \\
        0 & 0 & \sqrt{3} & 0
    \end{pmatrix}\,,
    J_z=\frac{1}{2}\begin{pmatrix}
        3 & 0 & 0 & 0 \\
        0 & 1 & 0 & 0 \\
        0 & 0 & -1 & 0 \\
        0 & 0 & 0 & -3
    \end{pmatrix}\,.
\end{equation}
}, $\vec{J}^3\equiv(J_x^3,J_y^3,J_z^3)$, $\mu_B$ the Bohr magneton, and $\kappa$, $q$ the isotropic and cubic Zeeman parameters. The action of $\vec{B}$ on the envelopes of the hole is accounted for by the substitution $\vec{p}\to-i\hbar\boldsymbol{\nabla}+e\vec{A}$ in $H_\mathrm{K}$, with $\vec{A}=(0,B_z x,B_x y-B_y x)$ the magnetic vector potential.

These equations are solved with a finite-difference (FD) discretization of the differential equations on a rectilinear mesh of the device, as done for example in Refs. \cite{martinez2022hole,Abadillo2023,Mauro2025b}.

\subsection{Effective Hamiltonian}

The FD solution of the LK equations provides accurate wave functions and energies at any bias and magnetic field. However, it is not suitable for fast explorations of the electronic structure of the DQD around a reference bias point $V^0\equiv(V_\mathrm{L}^0,V_\mathrm{R}^0,V_\mathrm{T}^0)$. We need, moreover, a tool to analyze the physical mechanisms (charge tunneling, SO coupling, ...) at work in the DQD.

For these purposes, we build an effective Hamiltonian $H_\mathrm{eff}$ in the minimal subspace spanned by the four first bonding/anti-bonding FD states $\{\ket{-\uparrow},\ket{-\downarrow},\ket{+\uparrow},\ket{+\downarrow}\}$ computed at $V=V^0$ and $\vec{B}=\vec{0}$. With $\delta V_i=V_i-V_i^0$, the (small) deviations of the gate voltages with respect to that bias point and $B_\alpha$ the components of the magnetic field along the $\{x,y,z\}$ axes, the $4\times 4$ matrix of this effective Hamiltonian can be expanded as \cite{MRodriguez2025}:
\begin{equation}
H_\mathrm{eff}=H_0+e\delta V_\mathrm{L} D_\mathrm{L}+e\delta V_\mathrm{R} D_\mathrm{R}+e\delta V_\mathrm{T} D_\mathrm{T}+\sum_{\alpha\in\{x,y,z\}}B_\alpha M_\alpha\,.
\label{eq:Heff}
\end{equation}
Here $H_0$ is the Hamiltonian at $V=V^0$ and $\vec{B}=\vec{0}$, $D_i$ is the matrix of the operator $d_i(\vec{r})=\partial V_t(\vec{r})/\partial V_i$ (the derivative of the total potential $V_t(\vec{r})$ with respect to the gate voltage $V_i$), and $M_\alpha$ is the matrix of the operator $\partial H_\mathrm{LK}/\partial B_\alpha$ (the derivative of the finite-difference LK Hamiltonian with respect to $B_\alpha$). $H_0=\mathrm{diag}(E_-,E_-,E_+,E_+)$ is thus diagonal by design ($E_-$ and $E_+$ being the spin-degenerate energies of the bonding and anti-bonding states at $V=V^0$). Moreover, $d_i(\vec{r})$ is nothing else than the potential created by a bias of 1\,V on gate $i\in\{{\rm L},{\rm R},{\rm T}\}$ with all other gates grounded, since the electrostatics is linear. Eq.~\eqref{eq:Heff} neglects $\propto B_\alpha^2$ corrections that are irrelevant in the linear response regime ($g$-matrix formalism \cite{Venitucci18}).

While this effective Hamiltonian is sufficient to compute energies and spin-photon couplings around the reference bias $V^0$, it does not lend itself to simple analysis and interpretation (because the states $\ket{\pm\sigma}$ mix left and right orbitals). We shall, therefore, transform this Hamiltonian to a suitably defined localized basis set $\{\ket{{\rm L}\Uparrow},\ket{{\rm L}\Downarrow},\ket{{\rm R}\Uparrow},\ket{{\rm R}\Downarrow}\}$, in order to extract the left/right $g$-matrices $g_\mathrm{L,R}$, the tunneling parameters $t_c$, $\theta_\mathrm{so}$ and $\vec{n}_\mathrm{so}$ (and the associated $g$-matrix $g_\mathrm{T}$ and vector $\vec{\mu}_\mathrm{T}$) of Eq.~(1) of the main text.

\subsection{The left and right orbitals}

To build this transformation, we first define a ``pure detuning'' axis in the $(\delta V_\mathrm{L},\delta V_\mathrm{R})$ plane of the stability diagram of the DQD. We look for the line $\delta V_\mathrm{L}=a_\mathrm{L}\delta V_d$, $\delta V_\mathrm{R}=-a_\mathrm{R}\delta V_d$ ($a_\mathrm{L}^2+a_\mathrm{R}^2=1$) of this plane such that the matrix $D_d=a_\mathrm{L} D_\mathrm{L}-a_\mathrm{R} D_\mathrm{R}$ has two opposite pairs of degenerate\footnote{The eigenvalues of $D_d$ are at least twofold degenerate owing to time-reversal symmetry constraints for an electric field operator (Kramers degeneracy).} eigenvalues $(-\alpha_d,-\alpha_d,+\alpha_d,+\alpha_d)$. We next choose $\{\ket{{\rm L}\Uparrow},\ket{{\rm L}\Downarrow}\}$ as the eigenvectors of $D_d$ with positive eigenvalue $+\alpha_d$, and $\{\ket{{\rm R}\Uparrow},\ket{{\rm R}\Downarrow}\}$ as the eigenvectors of $D_d$ with negative eigenvalue $-\alpha_d$. Indeed, $\delta V_d$ can then be interpreted as a pure detuning modulation as it symmetrically shifts the L and R orbital energies by an $\delta\varepsilon=\pm e\alpha_d\delta V_d$. The above condition on the eigenvalues is equivalent to $\Tr\,D_d=0$ (with $\Tr$ the trace), which leads to:
\begin{subequations}
\begin{align}
a_\mathrm{L}&=\pm\frac{\Tr\,D_\mathrm{R}}{\sqrt{\Tr^2\,D_\mathrm{L}+\Tr^2\,D_\mathrm{R}}} \\
a_\mathrm{R}&=\pm\frac{\Tr\,D_\mathrm{L}}{\sqrt{\Tr^2\,D_\mathrm{L}+\Tr^2\,D_\mathrm{R}}}\,.
\end{align}
\end{subequations}
We fix the signs so that $a_\mathrm{L}>0$. This construction is similar to maximally localized Wannier functions, using a detuning rather than a position operator. We call $U$ the rotation from the $\{\ket{{\rm L}\Uparrow},\ket{{\rm L}\Downarrow},\ket{{\rm R}\Uparrow},\ket{{\rm R}\Downarrow}\}$ to the $\{\ket{-\uparrow},\ket{-\downarrow},\ket{+\uparrow},\ket{+\downarrow}\}$ basis set.

In the $\{\ket{{\rm L}\Uparrow},\ket{{\rm L}\Downarrow},\ket{{\rm R}\Uparrow},\ket{{\rm R}\Downarrow}\}$ basis set, we can thus rewrite the Hamiltonian matrix as
\begin{equation}
H_\mathrm{eff}'=U^\dagger H_\mathrm{eff} U=H_0'+e\delta V_d D_d'+e\delta V_c D_c'+e\delta V_\mathrm{T} D_\mathrm{T}'+\sum_{\alpha\in\{x,y,z\}}B_\alpha M_\alpha'\,,
\label{eq:Heffprime}
\end{equation}
where $\delta V_d=a_\mathrm{L}\delta V_\mathrm{L}-a_\mathrm{R}\delta V_\mathrm{R}$ is the detuning voltage, $\delta V_c=a_\mathrm{R}\delta V_\mathrm{L}+a_\mathrm{L}\delta V_\mathrm{R}$ is the common mode voltage, $D_c'=a_\mathrm{R} D_\mathrm{L}'+a_\mathrm{L} D_\mathrm{R}'$ and $D_d'=\mathrm{diag}(\alpha_d,\alpha_d,-\alpha_d,-\alpha_d)$. We impose $\ket{{\rm L}\Uparrow}={\cal T}\ket{{\rm L}\Downarrow}$ and $\ket{{\rm R}\Uparrow}={\cal T}\ket{{\rm R}\Downarrow}$, with $\cal{T}$ the time-reversal symmetry operator. Therefore, $H_0'$ must be of the following form, imposed by time-reversal symmetry constraints:
\begin{equation}
H_0'=
\begin{pmatrix}
    \varepsilon_\mathrm{L} & 0 & t & t' \\
    0 & \varepsilon_\mathrm{L} & -t'^* & t^* \\
    t^* & -t' &\varepsilon_\mathrm{R} & 0 \\
    t'^* & t & 0 & \varepsilon_\mathrm{R}
\end{pmatrix}\,,
\label{eq:Heff2}
\end{equation}
while $D_c'$ and $D_\mathrm{T}'$ must be of the form:
\begin{equation}
D_i'=
\begin{pmatrix}
    \alpha_{i,{\rm L}} & 0 & \alpha_i & \alpha_i' \\
    0 & \alpha_{i,{\rm L}} & -\alpha_i'^* & \alpha_i^* \\
    \alpha_i^* & -\alpha_i' & \alpha_{i,{\rm R}} & 0 \\
    \alpha_i'^* & \alpha_i & 0 & \alpha_{i,{\rm R}}
\end{pmatrix}\,.
\end{equation}
We emphasize though that the transformation $U$ is defined up to arbitrary pseudo-spin rotations in the L and R subspaces (as the eigenvalues of $D_d$ are degenerate). The tunneling block of $H_0'$ (the matrix elements $t$ and $t'$) is thus largely arbitrary (with the only constraint $\sqrt{|t|^2+|t'|^2}=t_c$, the charge tunneling strength) unless a consistent definition of the pseudo-spins $\ket{\Uparrow}$ and $\ket{\Downarrow}$ has been chosen (see discussion about the choice of frame in section \ref{sec:frames}).

\subsection{Choosing the pseudo-spins}

Therefore, we choose the pseudo-spins $\ket{\Uparrow}$ and $\ket{\Downarrow}$ in the L and R dots so that the real part of the $j_z=+3/2$ envelope of $\ket{\Uparrow}$ is maximum (and, as a consequence of time-reversal symmetry, the real part of the $j_z=-3/2$ envelope of $\ket{\Downarrow}$ is maximum). That way, $\ket{\Uparrow}$ and $\ket{\Downarrow}$ best match the expectations of lowest-order perturbation theories\footnote{Given the form of the LK Hamiltonian [Eq.~\eqref{eq:LK}], $j_z=+3/2$ can not be mixed with $j_z=-3/2$ states to first order in the HH/LH coupling terms $R$ and $S$, and vice-versa.} for the HH/LH mixing in quantum dots \cite{Michal21,martinez2022hole,Abadillo2023,Marcellina17}, so that, for example, $t$ and $t'$ can be directly compared to the Rashba and Dresselhaus SO tunneling strengths given by these perturbation theories (see later discussions).

For that purpose, we proceed in two steps. Starting from arbitrary $\ket{{\rm L}\Uparrow}$ and $\ket{{\rm L}\Downarrow}$, we look for a new pseudo-spin
\begin{equation}
\ket*{{\rm L}\decorate{\Uparrow}}=\alpha\ket{{\rm L}\Uparrow}+\beta\ket{{\rm L}\Downarrow} \\
\end{equation}
such that the norm of the $j_z=-3/2$ envelope $\varphi_{-3/2}^{{\rm L}\decorate{\Uparrow}}(\vec{r})$ is minimum. We can express this norm as $\norm {\varphi_{-3/2}^{{\rm L}\decorate{\Uparrow}}}=\sqrt{\bra{\vec{u}}S\ket{\vec{u}}}$, where $\vec{u}=(\alpha,\beta)$ and $S$ is the positive definite overlap matrix:
\begin{equation}
S=
  \begin{pmatrix}
    \braket{\varphi_{-3/2}^{{\rm L}\Uparrow}}{\varphi_{-3/2}^{{\rm L}\Uparrow}} & \braket{\varphi_{-3/2}^{{\rm L}\Uparrow}}{\varphi_{-3/2}^{{\rm L}\Downarrow}} \\
    \braket{\varphi_{-3/2}^{{\rm L}\Downarrow}}{\varphi_{-3/2}^{{\rm L}\Uparrow}} & \braket{\varphi_{-3/2}^{{\rm L}\Downarrow}}{\varphi_{-3/2}^{{\rm L}\Downarrow}}
  \end{pmatrix}\,.
\end{equation}
The solution $\vec{u}$ that minimizes $\norm {\varphi_{-3/2}^{{\rm L}\decorate{\Uparrow}}}$ is thus the eigenvector of $S$ with minimal eigenvalue. Next, we look for a phase transformation $\ket*{{\rm L}\decorate{\Uparrow}'}=e^{i\gamma}\ket*{{\rm L}\decorate{\Uparrow}}$ such that the norm of the imaginary part of the $j_z=+3/2$ envelope $\varphi_{+3/2}^{{\rm L}\decorate{\Uparrow}'}(\vec{r})$ is minimum. Writing $e^{i\gamma}=a+ib$, $\norm{\Im\varphi_{+3/2}^{{\rm L}\decorate{\Uparrow}'}}=\sqrt{(\vec{v}|S'|\vec{v})}$ where $\vec{v}=(a,b)$ and $S'$ is the real symmetric matrix:
\begin{equation}
S'=
  \begin{pmatrix}
    (\Im\varphi_{+3/2}^{{\rm L}\decorate{\Uparrow}}|\Im\varphi_{+3/2}^{{\rm L}\decorate{\Uparrow}}) & (\Im\varphi_{+3/2}^{{\rm L}\decorate{\Uparrow}}|\Re\varphi_{+3/2}^{{\rm L}\decorate{\Uparrow}}) \\
    (\Re\varphi_{+3/2}^{{\rm L}\decorate{\Uparrow}}|\Im\varphi_{+3/2}^{{\rm L}\decorate{\Uparrow}}) & (\Re\varphi_{+3/2}^{{\rm L}\decorate{\Uparrow}}|\Re\varphi_{+3/2}^{{\rm L}\decorate{\Uparrow}})
  \end{pmatrix}\,.
\end{equation}
The solution $\vec{v}$ is thus, again, the eigenvector of $S'$ with smallest eigenvalue. The pseudo-spin $\ket*{{\rm L}\decorate{\Uparrow}'}$ is now defined up to a sign, which can be fixed by imposing that the real part of the $\varphi_{+3/2}$ envelope is essentially positive. The pseudo-spin $\ket*{{\rm L}\decorate{\Downarrow}'}$ follows from the time-reversal symmetry relation $\ket*{{\rm L}\decorate{\Downarrow}'}=-{\cal T}\ket*{{\rm L}\decorate{\Uparrow}'}$. The procedure is the same for the right spin orbitals $\ket*{{\rm R}\decorate{\Uparrow}'}$ and $\ket*{{\rm R}\decorate{\Downarrow}'}$. The $\{\ket*{{\rm L}\decorate{\Uparrow}'},\ket*{{\rm L}\decorate{\Downarrow}'},\ket*{{\rm R}\decorate{\Uparrow}'},\ket*{{\rm R}\decorate{\Downarrow}'}\}$ orbitals define the ``canonical'' basis set introduced in the main text. For simplicity, we drop the prime and bar on the spins hereafter; unless otherwise stated, $\{\ket{\rm L\Uparrow},\ket{\rm L\Downarrow},\ket{\rm R\Uparrow},\ket{\rm R\Downarrow}\}$ is thus the canonical basis set.

\subsection{Extraction of the parameters and calculation of the spin-photon coupling}

Once the canonical basis set has been defined, the $g$-matrices $g_\mathrm{L}$ and $g_\mathrm{R}$ of the left and right dots, the tunneling parameters $t$ and $t'$ (or equivalently $t_c$, $\theta_\mathrm{so}$ and $\vec{n}_\mathrm{so}$) and the associated $g$-matrix $g_\mathrm{T}$ and vector $\vec{\mu}_\mathrm{T}$ can be straightforwardly extracted by matching Eq.~\eqref{eq:Heffprime} with Eq.~(1) of the main text \cite{Venitucci18}. The L and R spin orbitals are constructed with the above procedure from the FD states $\{\ket{-\uparrow},\ket{-\downarrow},\ket{+\uparrow},\ket{+\downarrow}\}$ computed at $\vec{B}=\vec{0}$ and zero detuning between the dots (hence $\varepsilon_\mathrm{L}=\varepsilon_\mathrm{R}$ in $H_0^\prime$).

We introduce in the next section the Newton-Raphson algorithm used to find the gate voltages $V_\mathrm{L}$, $V_\mathrm{R}$ and $V_\mathrm{T}$ achieving zero detuning at given target tunnel coupling $t_c$ and chemical potential $\mu$.

We may use either Eq.~\eqref{eq:Heff} or Eq.~\eqref{eq:Heffprime} (which only differ by an unitary transform) for fast exploration of the spectrum of the DQD around a reference bias point $V^0\equiv(V_\mathrm{L}^0,V_\mathrm{R}^0,V_\mathrm{T}^0)$. The observables are reasonably accurate (with respect to full FD calculations) if the deviations $\delta V_i$ and the applied magnetic field are small enough. The spin-photon couplings $g_s$ reported in this work are actually computed with Eq.~(2) of the main text and the eigenstates of the four levels Hamiltonian \eqref{eq:Heff} (because we need a fast solver for the detailed angular maps of Fig.~2). We have checked the accuracy of the underlying $g$-matrix formalism by comparing with FD calculations of the spin-photon coupling at selected, finite magnetic fields.

\subsection{Tuning the devices}

Given side gate potentials $V_{\mathrm{S}_n}$, we tune $V_\mathrm{L}$, $V_\mathrm{R}$ and $V_\mathrm{T}$ in order to bring the device to zero detuning, target chemical $\varepsilon_\mathrm{L}=\varepsilon_\mathrm{R}=\mu$ and target charge tunnel coupling $t_t$. The choice of chemical potential $\mu$ essentially controls the size of the dots.

For that purpose, we start from a first guess bias $V_1\equiv(V_\mathrm{L},V_\mathrm{R},V_\mathrm{T})$ and compute the FD wave functions $\{\ket{-\uparrow},\ket{-\downarrow},\ket{+\uparrow},\ket{+\downarrow}\}$ at $\vec{B}=\vec{0}$. We then construct the effective Hamiltonian $H_\mathrm{eff}'$ at this bias point and extract the energies $\varepsilon_\mathrm{L}$, $\varepsilon_\mathrm{R}$ and tunnel coupling $t_c$. We next optimize the bias shifts $\delta V_\mathrm{L}$, $\delta V_\mathrm{R}$ and $\delta V_\mathrm{T}$ of this effective Hamiltonian with a Newton-Raphson algorithm, to best meet the conditions $\varepsilon_\mathrm{L}=\varepsilon_\mathrm{R}=\mu$ and $t_c=t_t$. We finally recompute the FD wave functions at bias $V_2\equiv(V_\mathrm{L}+\delta V_\mathrm{L},V_\mathrm{R}+\delta V_\mathrm{R},V_\mathrm{T}+\delta V_\mathrm{T})$ and iterate until convergence. The effective Hamiltonian $H_\mathrm{eff}'$ is accurate only for small enough steps $(\delta V_\mathrm{L},\delta V_\mathrm{R},\delta V_\mathrm{T})$; therefore, the steps taken during the first few iterations are pretty poor, but become better aimed when getting closer to the solution. Given the cost of FD calculations, this indirect optimization process is orders of magnitude faster than a direct Newton-Raphson with the FD Hamiltonian.

\section{Parameters of the Hamiltonian in the strained quantum well and bulk heterostructures}
\label{sec:parameters}

In this section, we give the parameters ($g$-matrices, tunnelings, ...) extracted in the devices of the main text and discuss additional bias configurations.

\subsection{Strained quantum well heterostructure}

\subsubsection{Hamiltonian of the strained heterostructure of Fig.~2a-d of the main text}
\label{sec:strainedQW}

\begin{figure*}[!t]
    \centering
    \includegraphics[width=\textwidth]{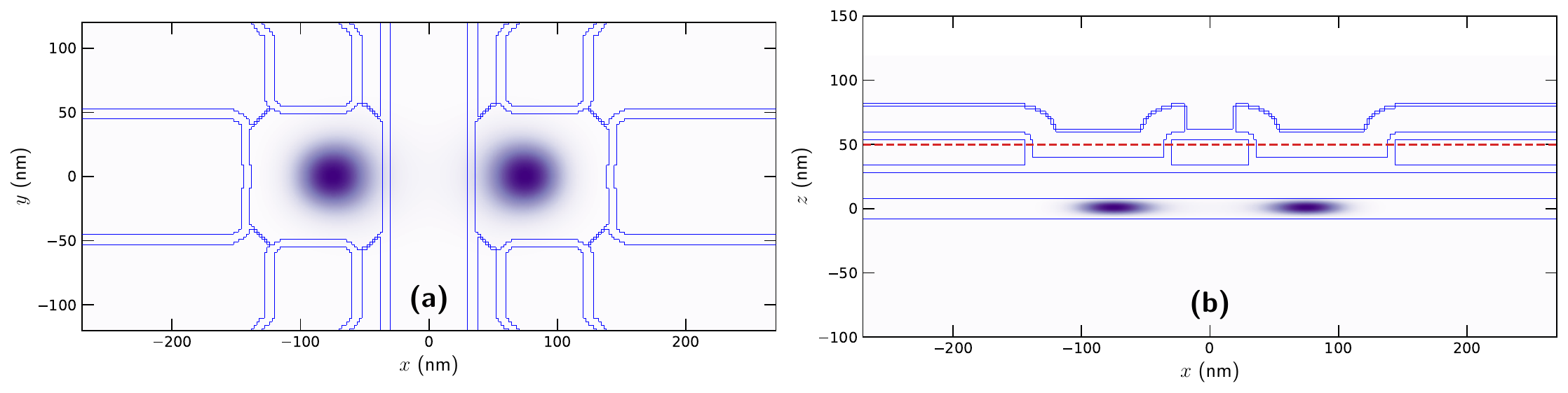}
    \caption{Map of the squared L and R orbitals of the strained quantum well in (a) the $(xy)$ plane at $z=0$, and (b) the $(xz)$ plane at $y=0$. The bias point (same as Fig.~2a-d of the main text) is $V_\mathrm{L}=V_\mathrm{R}=-20.5$\,mV and $V_\mathrm{T}=-5$\,mV (all side gates grounded). The blue lines outline the interfaces between the different materials; in panel (a) the materials are plotted in the plane shown by the dashed red line in (b) in order to highlight the position of the gates.}
    \label{fig:strainedsym}
\end{figure*}

We first consider the strained quantum well of Fig.~2a-d of the main text, at bias point $V_\mathrm{L}=V_\mathrm{R}=-20.5$\,mV and $V_\mathrm{T}=-5$\,mV (all side gates grounded). The L and R orbitals of this DQD, plotted in Fig.~\ref{fig:strainedsym}, are symmetric with respect to the $(yz)$ plane at $x=0$. The in-plane extensions of the dots are $\ell_x=\sqrt{\langle x^2\rangle-\langle x\rangle^2}=23.5$\,nm and $\ell_y=\sqrt{\langle y^2\rangle-\langle y\rangle^2}=20.4$\,nm. The HH/LH mixing is very small in this device ($\mathrm{total\ LH\ components}=0.15\%$) owing to the large HH/LH bandgap $\Delta$ opened by the structural confinement and biaxial strains $\varepsilon_{xx}=\varepsilon_{yy}=-0.61\%$, $\varepsilon_{zz}=0.45\%$ in the Ge well \cite{Mauro2025b}. Namely, 
\begin{equation}
\Delta=\Delta_c+\Delta_s\text{ with }\Delta_s=b_v(\varepsilon_{xx}+\varepsilon_{yy}-2\varepsilon_{zz})\,,
\end{equation}
where $\Delta_c\approx 25$\,meV is the contribution from vertical confinement in the 16-nm-thick Ge well and $\Delta_s=46$\,meV is the contribution from strains (with $b_v=-2.16$\,eV the uniaxial deformation potential of the valence band of Ge \cite{Fischetti96}).

\begin{figure*}[!t]
    \centering
    \includegraphics[width=0.575\textwidth]{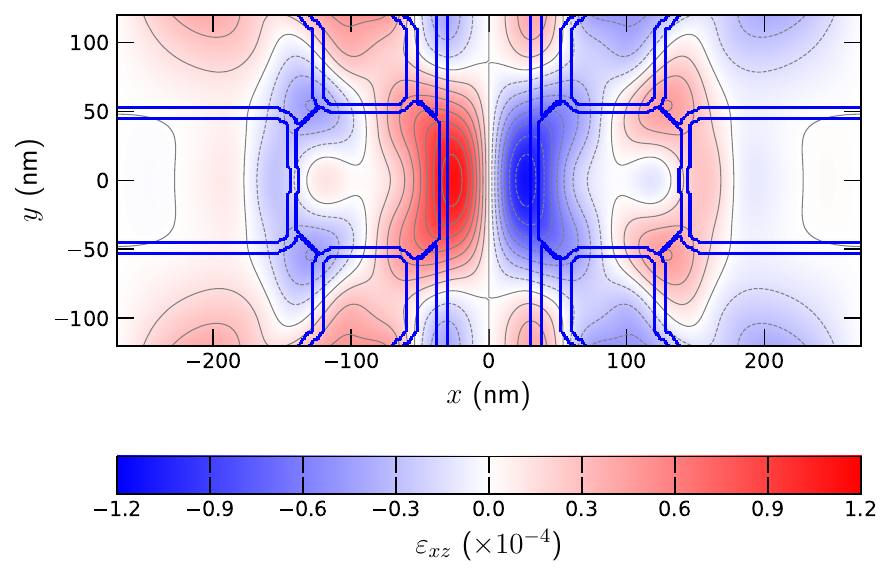}
    \caption{Map of the shear strains $\varepsilon_{xz}$ in the $(xy)$ plane of the bulk heterostructure at $z=0$. These shear strains are imprinted by the thermal contraction of the aluminium gates when the device is cooled down. Although small, they have a significant effect on the gyromagnetic axes of the hole \cite{Abadillo2023}. The blue lines outline the interfaces between the different materials in the same plane as in Fig.~\ref{fig:strainedsym}a, to highlight the position of the gates. The shear strains are almost the same in the strained quantum well as in the bulk heterostructure.}
    \label{fig:shearstrains}
\end{figure*}

The $g$-matrices and tunneling parameters extracted in the canonical basis set\footnote{The squared norms of the $(\varphi_{+3/2},\varphi_{+1/2},\varphi_{-1/2},\varphi_{-3/2})$ envelopes [Eq.~\eqref{eq:envelopes}] of the L and R $\ket{+\tfrac{3}{2}}$ states are $(99.84\%,0.15\%,0.01\%,0.00\%)$. Those of the time-reversal symmetric $\ket{-\tfrac{3}{2}}$ states are $(0.00\%,0.01\%,0.15\%,99.84\%)$.} with the methodology of section~\ref{sec:methodology} are:
\begin{gather}
    g_\mathrm{L}=
    \begin{pmatrix}
        0.234 & 0 & 0.004 \\
        0 & -0.046 & 0 \\
        0.048 & 0 & 13.78 
    \end{pmatrix},\,
    g_\mathrm{R}=
    \begin{pmatrix}    
        0.234 & 0 & -0.004 \\
        0 & -0.046 & 0 \\
        -0.048 & 0 & 13.78
    \end{pmatrix}; \nonumber \\
    g_\mathrm{T}=
    \begin{pmatrix}
        0.013 & 0 & 0 \\
        0 & 0.006 & 0 \\
        0 & 0 & 0.032 
    \end{pmatrix},\,
    \vec{\mu}_\mathrm{T}=
    \begin{pmatrix}
        0 \\
        -0.479 \\
        0 
    \end{pmatrix}\frac{\mu\mathrm{eV}}{\mathrm{T}}; \nonumber \\
    t_c/h=3.5\,\mathrm{GHz},\,\theta_\mathrm{so}=0.83\degree,\,\vec{n}_\mathrm{so}=\vec{y}.
    \label{eq:strainedQW}
\end{gather}
The tunneling angle $\theta_\mathrm{so}$ is very small and practically negligible. As discussed in section~\ref{sec:SOC}, Rashba SO interactions are, indeed, strongly limited by the small HH/LH mixing in strained germanium. The shape of the $g$-matrices is consistent with the symmetries of the system \cite{Venitucci18} and with perturbation theories for the HH/LH mixing \cite{Michal21,martinez2022hole,Abadillo2023}. The large gyromagnetic anisotropy ($g_{zz}\gg g_{xx},\,|g_{yy}|$ in both dots) is typical of almost pure heavy-holes. Surprisingly, the in-plane $g$-factors $g_{xx}>|g_{yy}|$ are representative \cite{Mauro24} of dots elongated along $x$ (yet $\ell_x\approx\ell_y$). This is a fingerprint of the significant one-dimensional character of the DQD (and, in particular, of the more delocalized LH envelopes the HH ground-state is coupling to). The off-diagonal $g_{zx}$ term (opposite in the two dots) results from the coupling\footnote{Namely, from the interplay \cite{martinez2022hole,Abadillo2023} between the $S_{\mathrm{K}/\varepsilon}$ terms of Eq.~\eqref{eq:LK} and the Zeeman Hamiltonian $H_\mathrm{Z}$ [Eq.~\eqref{eq:Zeeman}].} between the in- and out-of-plane motions of the hole \cite{martinez2022hole} and from the small shear strains $\varepsilon_{xz}$ imprinted by the thermal contraction of the metal gates upon cool-down (see Fig.~\ref{fig:shearstrains}) \cite{Abadillo2023}. To proceed further with the analysis, we can factorize the $g$-matrices of the dots as \cite{Venitucci18}
\begin{equation}
g=Ug_dV^T\,,
\label{eq:svd}
\end{equation}
where $g_d=\mathrm{diag}(g_1,g_2,g_3)$ is the diagonal matrix of principal $g$-factors, the columns $(\vec{v}_1,\vec{v}_2,\vec{v}_3)$ of $V$ are the principal magnetic axes, and the columns $(\vec{u}_1,\vec{u}_2,\vec{u}_3)$ of $U$ are the principal spin axes. With the magnetic field $\vec{B}=(B_1,B_2,B_3)$ expressed in the principal magnetic axes set, and the spin quantized along $\vec{u}_3$,\footnote{More precisely, the new pseudo-spin basis set (the columns of a $2\times2$ unitary transform $R$) must be chosen so that the matrix representations of $R^\dagger(\vec{u}_i\cdot\vec{\sigma})R$ are the usual Pauli matrices $\sigma_1\equiv\sigma_x$, $\sigma_2\equiv\sigma_y$, $\sigma_3\equiv\sigma_z$.} the effective Zeeman Hamiltonian of the hole simply becomes:
\begin{equation}
H_\mathrm{Zeeman}=\frac{1}{2}\mu_B\left(g_1B_1\sigma_x+g_2B_2\sigma_y+g_3B_3\sigma_z\right)\,.
\end{equation}
In the present case, $g_1\approx g_{xx}=0.234$, $g_2\approx g_{yy}=-0.046$, $g_3\approx g_{zz}=13.78$ in both dots, and $U\approx\mathbb{1}_3$ (the $3\times 3$ identity matrix). The vectors $(\vec{v}_1,\vec{v}_2,\vec{v}_3)$ are the $(\vec{x},\vec{y},\vec{z})$ axes rotated around $\vec{y}$ by $\delta\theta_\mathrm{L}^m\approx g_{zx}/g_{zz}=0.2\degree$ in the L dot, and by $\delta\theta_\mathrm{R}^m=-\delta\theta_\mathrm{L}^m$ in the R dot. $g_{zx}$ thus essentially leaves the principal spin axes invariant but symmetrically tilts the principal magnetic axes of the L and R dots. A magnetic field along $\vec{x}$ hence rocks from one side of the $(\vec{v}_1,\vec{v}_2)$ plane to the other when the hole tunnels between the dots. Although $\delta\theta_\mathrm{L/R}^m$ is small, the effects of these excursions on the pseudo-spin precession axes $\vec{\omega}_\mathrm{L}=\mu_Bg_\mathrm{L}\vec{B}/\hbar$ and $\vec{\omega}_\mathrm{R}=\mu_Bg_\mathrm{R}\vec{B}/\hbar$ are amplified by the large $g_{zz}/g_{xx}$ ratio, which can give rise to a significant spin-flip tunneling strength $t_\mathrm{sf}$. Indeed, as discussed in the main text and shown in section~\ref{sec:frames}, 
\begin{equation}
t_\mathrm{sf}=t_c\sin\left(\frac{\Theta_\mathrm{LR}}{2}\right)=t_c\sqrt{\frac{1-\cos\Theta_\mathrm{LR}}{2}}
\label{eq:tsfSM}
\end{equation}
when $\theta_\mathrm{so}$ is negligible, with $\Theta_\mathrm{LR}$ the angle between the vectors $\vec{\omega}_\mathrm{L}$ and $\vec{\omega}_\mathrm{R}$. Thus
\begin{equation}
\cos\Theta_\mathrm{LR}\approx\frac{g_{xx}^2-g_{zx}^2}{g_{xx}^2+g_{zx}^2}
\label{eq:thetalr}
\end{equation}
when $\vec{B}\parallel\vec{x}$, which yields:
\begin{equation}
t_\mathrm{sf}=t_c\frac{|g_{zx}|}{\sqrt{g_{xx}^2+g_{zx}^2}}\,.
\end{equation}
The spin-flip tunneling matrix element is thus at best $t_\mathrm{sf}\approx 0.2t_c$ as evidenced by Fig.~2c of the main text. Eq.~\eqref{eq:tsfSM} also nicely explains other features of that figure. First, $\cos\Theta_\mathrm{LR}=1$ (thus $t_\mathrm{sf}=0$ and $g_s=0$) when $\vec{B}\parallel\vec{y}$ because $g_\mathrm{L}\vec{y}\parallel g_\mathrm{R}\vec{y}\parallel\vec{y}$ (the rotations of the principal magnetic axes leave $\vec{y}$ invariant). Moreover, $\cos\Theta_\mathrm{LR}$ rapidly tends to 1 once the magnetic field goes out-of-plane because $g_\mathrm{L}\vec{B}$ and $g_\mathrm{R}\vec{B}$ both lock onto the $z$ axis owing to the very large $g_{zz}/g_{xx}$ ratio. The full width at half maximum of the in-plane peak of Fig.~2 (main text) is actually $\Delta\theta\approx 2g_{xx}/g_{zz}$ at $\varphi=0$.

\subsection{Unstrained bulk heterostructure}

\subsubsection{Hamiltonian of the bulk heterostructure of Fig.~2e-h of the main text}
\label{sec:symmetricsqueezed}

\begin{figure*}[!t]
    \centering
    \includegraphics[width=\textwidth]{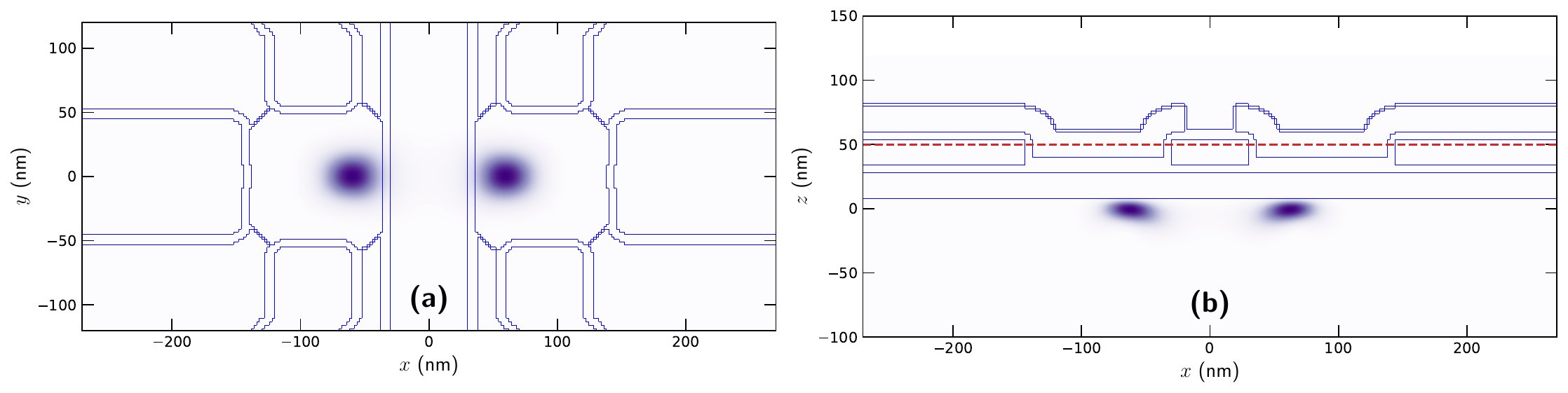}
    \caption{Map of the squared L and R orbitals of the bulk heterostructure in (a) the $(xy)$ plane at $z=0$, and (b) the $(xz)$ plane at $y=0$. The bias point (same as Fig.~2e-h of the main text) is $V_\mathrm{L}=V_\mathrm{R}=-43.4$\,mV, $V_\mathrm{T}=-11.6$\,mV, $V_{\mathrm{S}_1}=V_{\mathrm{S}_2}=V_{\mathrm{S}_3}=V_{\mathrm{S}_4}=30$\,mV. The blue lines outline the interfaces between the different materials; in panel (a) the materials are plotted in the plane shown by the dashed red line in (b) in order to highlight the position of the gates.}
    \label{fig:bulksqueezedsym}
\end{figure*}

We next consider the bulk heterostructure of Fig.~2e-h of the main text, at bias point $V_\mathrm{L}=V_\mathrm{R}=-43.4$\,mV and $V_\mathrm{T}=-11.6$\,mV (top and bottom side gates bias $V_{\mathrm{S}_1}=V_{\mathrm{S}_2}=V_{\mathrm{S}_3}=V_{\mathrm{S}_4}=30$\,mV). The L and R orbitals of this DQD, plotted in Fig.~\ref{fig:bulksqueezedsym}, are symmetric with respect to the $(yz)$ plane at $x=0$. The dots are slightly squeezed along $y$ ($\ell_x=18.7$\,nm and $\ell_y=13.8$\,nm) by the bias applied to the side gates. This squeezing reduces HH/LH mixing, but actually enhances the spin-photon coupling and makes it more robust against bias fluctuations (see the discussion of quasi-circular dots in section~\ref{sec:symmetriccircular}).

The $g$-matrices and tunneling parameters extracted in the canonical basis set\footnote{The squared norms of the $(\varphi_{+3/2},\varphi_{+1/2},\varphi_{-1/2},\varphi_{-3/2})$ envelopes [Eq.~\eqref{eq:envelopes}] of the L and R $\ket{+\tfrac{3}{2}}$ states are $(78.37\%,14.16\%,4.86\%,2.61\%)$. Those of the time-reversal symmetric $\ket{-\tfrac{3}{2}}$ states are $(2.61\%,4.86\%,14.16\%,78.37\%)$.} are:
\begin{gather}
    g_\mathrm{L}=
    \begin{pmatrix}
        0.454 & 0     & 0.571 \\
        0     & 0.694 & 0 \\
        1.180 & 0     & 1.086 
    \end{pmatrix},\,
    g_\mathrm{R}=
    \begin{pmatrix}
        0.454 & 0     & -0.571 \\
        0     & 0.694 & 0 \\
       -1.180 & 0     & 1.086 
    \end{pmatrix}; \nonumber \\
    g_\mathrm{T}=
    \begin{pmatrix}
        0.131 & 0 & 0 \\
        0 & 0.233 & 0 \\
        0 & 0 & 0.278 
    \end{pmatrix},\,
    \vec{\mu}_\mathrm{T}=
    \begin{pmatrix}
        0 \\
        -8.228 \\
        0 
    \end{pmatrix}\frac{\mu\mathrm{eV}}{\mathrm{T}}; \nonumber \\
    t_c/h=3.5\,\mathrm{GHz},\,\theta_\mathrm{so}=60.77\degree,\,\vec{n}_\mathrm{so}=\vec{y}. \label{eq:bulkparams}
\end{gather}
With respect to the strained quantum well [Eq.~\eqref{eq:strainedQW}], the (diagonal) gyromagnetic anisotropy is much reduced by the stronger HH/LH mixing ($\mathrm{LH}=15\%$), but the off-diagonal elements $g_{zx}$ and $g_{xz}$ of $g_\mathrm{L}$ and $g_\mathrm{R}$ are larger. The principal gyromagnetic factors and axes of $g_\mathrm{L}$ are:
\begin{equation}
    g_d=
    \begin{pmatrix}
        -0.103 & 0 & 0 \\
        0 & 0.694 & 0 \\
        0 & 0 & 1.759
    \end{pmatrix},\,
    V=
    \begin{pmatrix}
        0.696 & 0 &  0.718 \\
        0     & 1 &  0 \\
       -0.718 & 0 &  0.696 
    \end{pmatrix},\,
    U=
    \begin{pmatrix}
         0.911 & 0 & 0.411 \\
         0     & 1 & 0 \\
        -0.411 & 0 & 0.911
    \end{pmatrix}\,,
\end{equation}
and those of $g_\mathrm{R}$ are:
\begin{equation}
    g_d=
    \begin{pmatrix}
        -0.103 & 0 & 0 \\
        0 & 0.694 & 0 \\
        0 & 0 & 1.759
    \end{pmatrix},\,
    V=
    \begin{pmatrix}
        0.696 & 0 &  -0.718 \\
        0     & 1 &  0 \\
        0.718 & 0 &  0.696 
    \end{pmatrix},\,
    U=
    \begin{pmatrix}
         0.911 & 0 & -0.411 \\
         0     & 1 & 0 \\
         0.411 & 0 & 0.911
    \end{pmatrix}\,.
\end{equation}
Strikingly, the rotations undergone by the principal magnetic axes (the columns of $V$) are very large in this heterostructure ($\delta\theta_\mathrm{L}^m=-\delta\theta_\mathrm{R}^m=45.9\degree$). The principal pseudo-spin axes (the columns of $U$) remain much closer to the crystallographic $\{\vec{x},\vec{y},\vec{z}\}$ axes (they are rotated around $\vec{y}$ by a smaller angle $\delta\theta_\mathrm{L}^s=-\delta\theta_\mathrm{R}^s=24.3\degree$). These rotations are promoted by the ``tilt'' of the L and R orbitals induced by the interplay between the potential of the different gates (see Fig.~\ref{fig:bulksqueezedsym} and Fig.~3 of the main text), and by the small inhomogeneous strains imprinted by the thermal contraction of the metal gates (Fig.~\ref{fig:shearstrains}) \cite{Abadillo2023}. They are actually eased by the lower gyromagnetic anisotropy, which otherwise tends to lock $\vec{u}_3$ and $\vec{v}_3$ onto $z=[001]$. Discarding shear strains in the LK Hamiltonian [setting $R_\varepsilon=S_\varepsilon=0$ in Eqs.~\eqref{eq:PQRSstrains}] reduces $\delta\theta_\mathrm{R}^m=-\delta\theta_\mathrm{L}^m=27.7\degree$ and $\delta\theta_\mathrm{R}^s=-\delta\theta_\mathrm{L}^s=-12.6\degree$ but leaves $\theta_\mathrm{so}=59.5\degree$ almost invariant.

The rotations of the gyromagnetic axes as well as the large $\theta_\mathrm{so}=60.77\degree$ are indicative of a general enhancement of SO coupling by the greater HH/LH mixing. We discuss in more details in section~\ref{sec:SOC} the nature of the Rashba-like interactions giving rise to this sizable $\theta_\mathrm{so}$. The magnetotunneling terms $g_\mathrm{T}$ and $\vec{\mu}_\mathrm{T}$ (which imply a steeper dependence of spin tunneling on the magnetic field amplitude) are also significantly strengthened by the enhanced SO coupling \cite{rodriguezmena2025}. Neglecting the magnetotunneling Hamiltonian $H_\mathrm{MT}$ actually results in a $\approx 20\%$ error on the net spin-photon coupling $g_s$ computed for $\vec{B}\parallel\vec{x}$ and $\omega_s/2\pi=2$\,GHz. We provide a detailed analysis of the role of magnetotunneling in this device in section \ref{sec:frames}.

Shall we discard $\theta_\mathrm{so}$ and apply Eq.~\eqref{eq:thetalr} with the above $g_{xx}=0.454$ and $g_{zx}=\pm 1.180$, we would find a near optimal $\Theta_\mathrm{LR}=137.91\degree$ when $\vec{B}\parallel\vec{x}$ ($\cos\Theta_\mathrm{LR}=-0.74$, $t_\mathrm{sf}=0.93t_c$). Therefore, the tunneling Hamiltonian $H_\mathrm{T}$ cancels part of the spin rotation brought by the mismatch between $g_\mathrm{L}$ and $g_\mathrm{R}$ (as $t_\mathrm{sf}$ is practically smaller). To understand why, we show in section \ref{sec:frames} that $\theta_\mathrm{so}$ can be eliminated \cite{Geyer24} by a transformation $U_\mathrm{so}^\mathrm{L/R}=\exp(\mp i\theta_\mathrm{so}\vec{n}_\mathrm{so}\cdot\vec{\sigma}/2)$ of the L and R pseudo-spin subspaces, which block-diagonalizes $H_\mathrm{T}\to t_c\sigma_0\tau_x$ but rotates the $g$-matrices $g_\mathrm{L}\to\hat{g}_\mathrm{L}={\cal R}(-\theta_\mathrm{so},\vec{n}_\mathrm{so})g_\mathrm{L}$ and $g_\mathrm{R}\to\hat{g}_\mathrm{R}={\cal R}(\theta_\mathrm{so},\vec{n}_\mathrm{so})g_\mathrm{R}$ [with ${\cal R}(\theta,\vec{n})$ the real space rotation of angle $\theta$ around $\vec{n}$]. The matrices $\hat{g}_\mathrm{L/R}$ in this ``SO frame'' have the same principal $g$-factors and magnetic axes (which are observables) as $g_\mathrm{L/R}$, but different principal spin axes [$U\to\hat{U}={\cal R}(\pm\theta_\mathrm{so},\vec{n}_\mathrm{so})U$]. Equation \eqref{eq:tsf} still holds in this frame, with ${\Theta}_\mathrm{LR}$ replaced by the angle $\widehat{\Theta}_\mathrm{LR}$ between the vectors $\hat{\vec{\omega}}_\mathrm{L}=\mu_B\hat{g}_\mathrm{L}\vec{B}/\hbar$ and $\hat{\vec{\omega}}_\mathrm{R}=\mu_B\hat{g}_\mathrm{R}\vec{B}/\hbar$. For $\vec{B}\parallel\vec{x}$, this angle is simply
\begin{equation}
\widehat{\Theta}_\mathrm{LR}=\Theta_\mathrm{LR}+2\theta_\mathrm{so}=259.45\degree
\end{equation}
so that $t_\mathrm{sf}=t_c\mathrm{sin}(\widehat{\Theta}_\mathrm{LR}/2)=0.77t_c$ is indeed smaller than expected from the sole mismatch between the canonical $g$-matrices $g_\mathrm{L}$ and $g_\mathrm{R}$. In other words, the SO rotation by $2\theta_\mathrm{so}$ brings the slightly under-optimal $\Theta_\mathrm{LR}=137.91\degree$ far beyond $\widehat{\Theta}_\mathrm{LR}=180\degree$. This suggests that the characteristic SO interaction lengths in this device are actually shorter than inter-dot distance $s=170$\,nm. We will come back to this argument in section~\ref{sec:SOC}.

\subsubsection{Dependence on confinement}
\label{sec:symmetriccircular}

\begin{figure*}[!t]
    \centering
    \includegraphics[width=\textwidth]{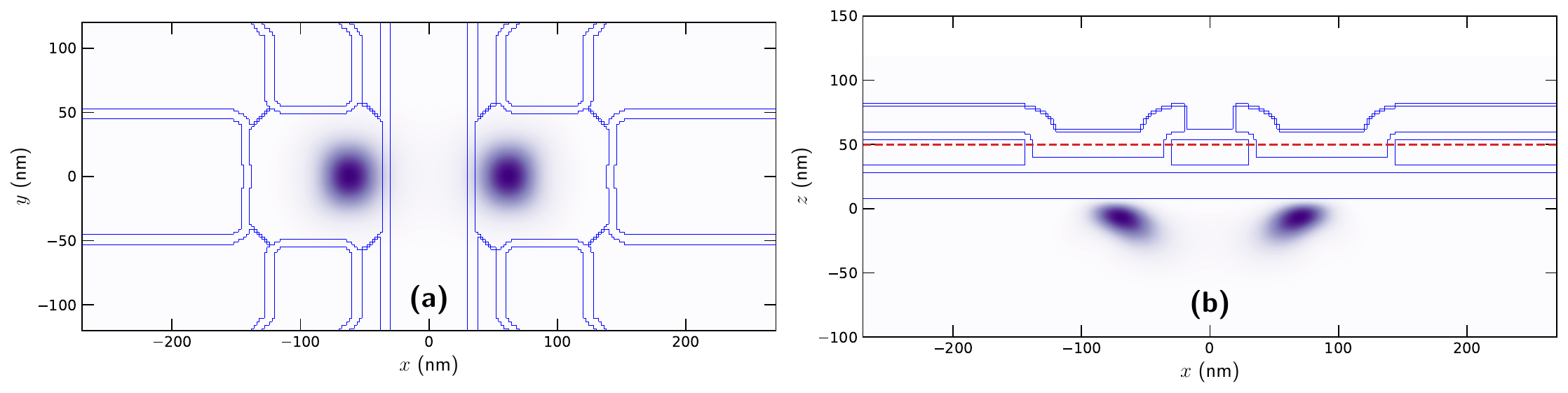}
    \caption{Map of the squared L and R orbitals of the bulk heterostructure in (a) the $(xy)$ plane at $z=0$, and (b) the $(xz)$ plane at $y=0$. The bias point is $V_\mathrm{L}=V_\mathrm{R}=-20.5$\,mV and $V_\mathrm{T}=-4.7$\,mV with all side gates grounded. The blue lines outline the interfaces between the different materials; in panel (a) the materials are plotted in the plane shown by the dashed red line in (b) in order to highlight the position of the gates.}
    \label{fig:bulksym}
\end{figure*}

We can further support this finding by looking at the dependence of the spin-photon coupling on the plunger gate bias. As the device of section~\ref{sec:symmetricsqueezed} is in a saturation regime where all relevant quantities (HH/LH mixing, $g_s$, $t_\mathrm{sf}$...) are little dependent on the plunger gate voltages (at fixed $t_c$), we start from a different bias point ($V_\mathrm{L}=V_\mathrm{R}=-20.5$\,mV and $V_\mathrm{T}=-4.7$\,mV with all side gates grounded) where the dots are less confined and more circular ($\ell_x=23.2$\,nm and $\ell_y=21.2$\,nm). The L and R orbitals of this DQD are plotted in Fig.~\ref{fig:bulksym}.

\begin{figure*}[!t]
    \centering
    \includegraphics[width=\textwidth]{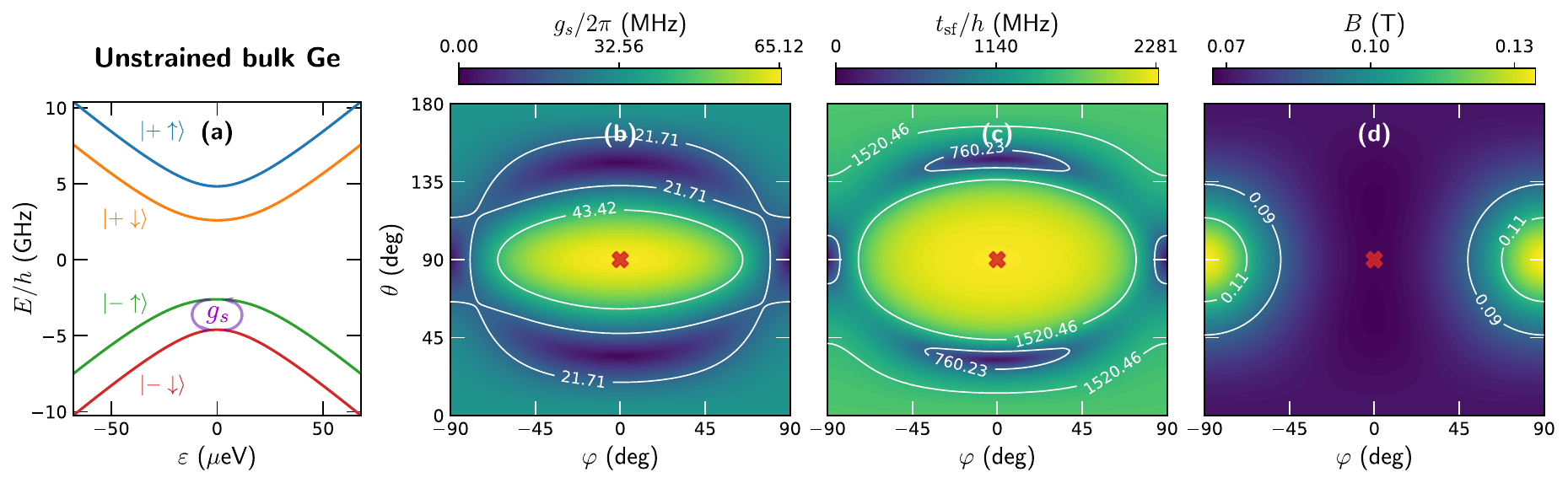}
    \caption{Spin-photon coupling in the bulk Ge heterostructure ($V_\mathrm{L}=V_\mathrm{R}=-20.5$\,mV and $V_\mathrm{T}=-4.7$\,mV with all side gates grounded): (a) Spectrum of the DQD as a function of detuning energy $\varepsilon$ ($\vec{B}\parallel\vec{x}$). The resonator couples to the hole in the $\ket{-\uparrow}$ and $\ket{-\downarrow}$ states; (b) Spin-photon coupling $g_s$, (c) spin-flip tunnel coupling $t_\mathrm{sf}$, (d) magnetic field amplitude at resonance ($\omega_s/2\pi=2$\,GHz) as a function of the orientation of $\vec{B}$. The red cross highlights the optimal orientation $\vec{B}\parallel\vec{x}$.}
    \label{fig:gsbulksym}
\end{figure*}

The Hamiltonian parameters at this bias point are:
\begin{gather}
    g_\mathrm{L}=
    \begin{pmatrix}
        1.405 & 0     & 1.578 \\
        0     & 0.993 & 0 \\
        2.674 & 0     & 1.349 
    \end{pmatrix},\,
    g_\mathrm{R}=
    \begin{pmatrix}
        1.405 & 0     & -1.578 \\
        0     & 0.993 & 0 \\
        -2.674 & 0     & 1.349 
    \end{pmatrix}; \nonumber \\
    g_\mathrm{T}=
    \begin{pmatrix}
        -0.122 & 0 & 0 \\
        0 & -0.890 & 0 \\
        0 & 0 & -0.202 
    \end{pmatrix},\,
    \vec{\mu}_\mathrm{T}=
    \begin{pmatrix}
        0 \\
        7.083 \\
        0 
    \end{pmatrix}\frac{\mu\mathrm{eV}}{\mathrm{T}}; \nonumber \\
    t_c/h=3.5\,\mathrm{GHz},\,\theta_\mathrm{so}=77.41\degree,\,\vec{n}_\mathrm{so}=\vec{y}.
\end{gather}

The spin-photon coupling $g_s$ at zero detuning, the spin-flip tunnel matrix element $t_\mathrm{sf}$ and the magnetic field amplitude at resonance ($\omega_s/2\pi=2$\,GHz) are plotted as a function of the orientation of $\vec{B}$ in Fig.~\ref{fig:gsbulksym} (along with the hole spectrum as a function of detuning). The maximum $g_s/2\pi=65.12$\,MHz and $t_\mathrm{sf}=0.65t_c$ ($\vec{B}\parallel\vec{x}$) remain pretty poor despite a strong HH/LH mixing ($\mathrm{LH}=38.3\%$) and SO angle $\theta_\mathrm{so}=77.41\degree$. As a matter of fact, the same analysis as in the previous section yields $\Theta_\mathrm{LR}=124.56\degree$ from the mismatch of the canonical $g$-matrices but $\widehat{\Theta}_\mathrm{LR}=\Theta_\mathrm{LR}+2\theta_\mathrm{so}=279.38\degree$ once the effects of SO tunneling are included.

\begin{figure*}[!t]
    \centering
    \includegraphics[width=0.35\textwidth]{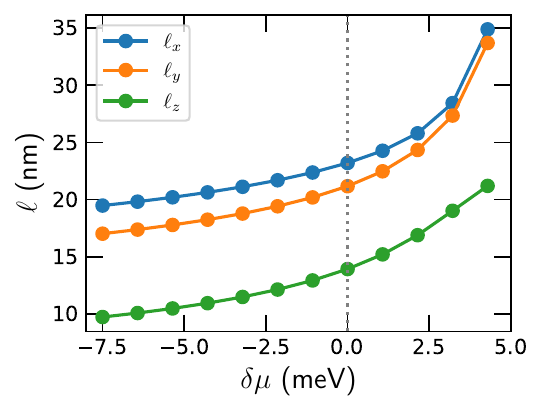}
    \caption{Sizes $\ell_x$, $\ell_y$ and $\ell_z$ of the dots in the bulk heterostructure as a function of the plunger gate voltages, characterized by the chemical potential shift $\delta\mu$ with respect to $V_\mathrm{L}=V_\mathrm{R}=-20.5$\,mV (all side gates grounded). The voltage $V_\mathrm{T}$ is adjusted so that $t_c/h=3.5$\,GHz.}
    \label{fig:lsvsEfield}
\end{figure*}

\begin{figure*}[!t]
    \centering
    \includegraphics[width=0.7\textwidth]{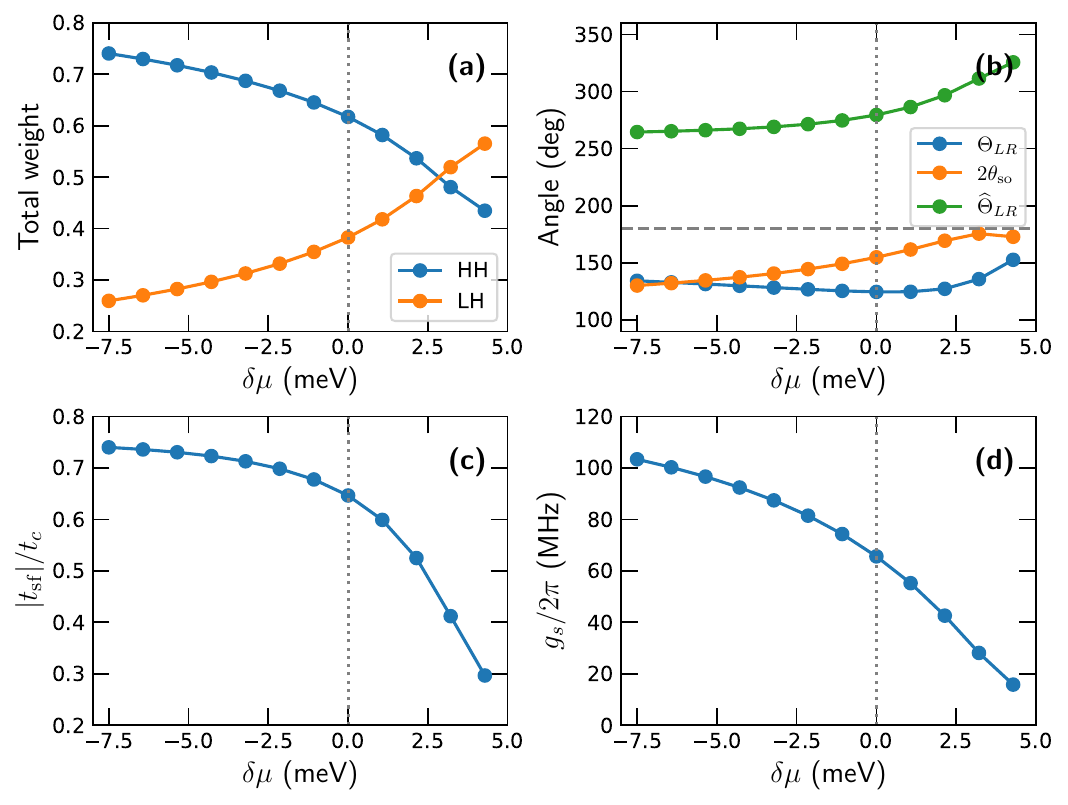}
    \caption{(a) HH/LH mixing in the bulk heterostructure as a function of the plunger gate voltages, characterized by the chemical potential shift $\delta\mu$ with respect to $V_\mathrm{L}=V_\mathrm{R}=-20.5$\,mV (all side gates grounded). The voltage $V_\mathrm{T}$ is adjusted so that $t_c/h=3.5$\,GHz. (b) Angles $2\theta_\mathrm{so}$, $\Theta_\mathrm{LR}$ and $\widehat{\Theta}_\mathrm{LR}$ as a function of $\delta\mu$ ($\vec{B}\parallel\vec{x}$). The horizontal, dashed line is the optimal $\widehat{\Theta}_\mathrm{LR}=180\degree$. (c) Spin-flip tunnel matrix element $t_\mathrm{sf}$ as a function of $\delta\mu$ ($\vec{B}\parallel\vec{x}$). (d) Spin-photon coupling $g_s$ as a function of $\delta\mu$ ($\vec{B}\parallel\vec{x}$ and $\omega_s/2\pi=2$\,GHz).}
    \label{fig:gsvsEfield}
\end{figure*}

We can bias the plunger gates more negatively in order to squeeze the dots at the Ge/Ge$_{0.8}$Si$_{0.2}$ interface (while acting on $V_\mathrm{T}$ to keep $t_c/h=3.5$\,GHz constant). This is expected to increase the HH/LH bandgap, and decrease the HH/LH mixing and $\theta_\mathrm{so}$. The size of the dots is plotted in Fig.~\ref{fig:lsvsEfield} as a function of the chemical potential shift $\delta\mu$ (measured with respect to the original bias point $V_\mathrm{L}=V_\mathrm{R}=-20.5$\,mV). The HH/LH mixing, the angles $\theta_\mathrm{so}$, $\Theta_\mathrm{LR}$ and $\widehat{\Theta}_\mathrm{LR}$, as well as $t_\mathrm{sf}$ and the spin-photon coupling $g_s$ ($\vec{B}\parallel\vec{x}$) are plotted in Fig.~\ref{fig:gsvsEfield}.

As expected, the dots get more confined (both vertically and laterally) when the electric field between the gates increases ($\delta\mu<0$) \cite{Mauro2025b}. Moreover, the HH/LH mixing decreases, as does $\theta_\mathrm{so}$, and, to a smaller extent, $\Theta_\mathrm{LR}$ (because the dots move to larger shear strains). This is consistent with a weakening of SO coupling effects. However, the net $\widehat{\Theta}_\mathrm{LR}$ gets closer to $180\degree$, so that both $t_\mathrm{sf}$ and $g_s$ actually increase (at least in the investigated bias range). Note that $g_s$ grows faster than $t_\mathrm{sf}$ due to favorable magnetotunneling effects and higher-order corrections beyond Eq.~(3) of the main text. The optimization of the spin-photon coupling in bulk heterostructures therefore calls for a careful control of the confinement. This provides, on the other hand, opportunities for fine tuning (as done by squeezing in section~\ref{sec:symmetricsqueezed}) and for device engineering. As an illustration, we discuss in section~\ref{sec:insertion} how the insertion of a thin GeSi layer below the dots allows for adjusting the SO interactions.

\subsubsection{Effects of dots asymmetry}

\begin{figure*}[!t]
    \centering
    \includegraphics[width=\textwidth]{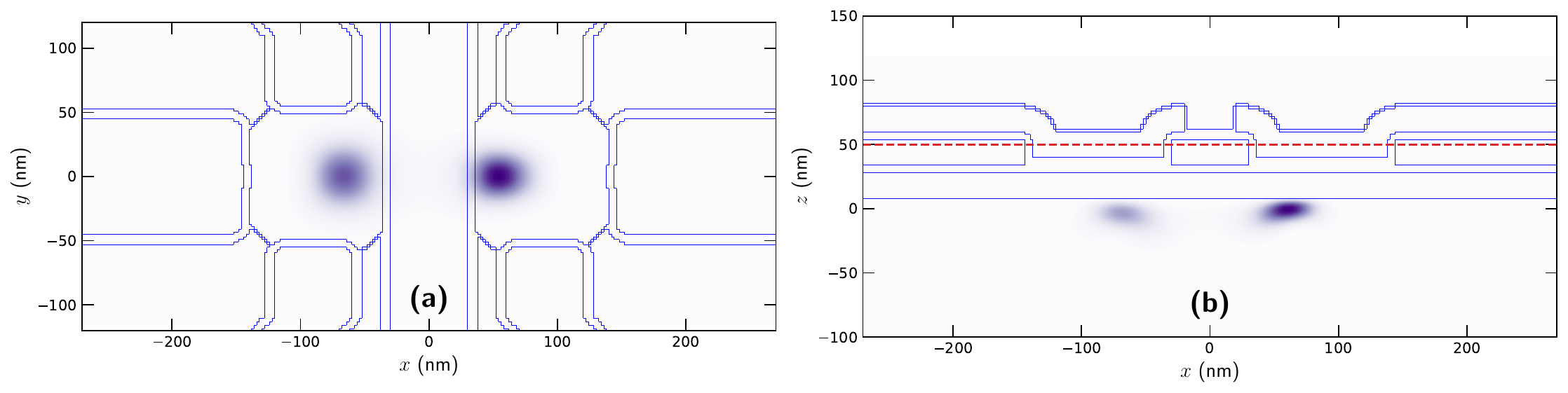}
    \caption{Map of the squared L and R orbitals of the bulk heterostructure in (a) the $(xy)$ plane at $z=0$, and (b) the $(xz)$ plane at $y=0$. The bias point is $V_\mathrm{L}=-24.2$\,mV, $V_\mathrm{R}=-46.9$\,mV, $V_\mathrm{T}=-9.8$\,mV, $V_{\mathrm{S}_2}=V_{\mathrm{S}_4}=40$\,meV. The blue lines outline the interfaces between the different materials; in panel (a) the materials are plotted in the plane shown by the dashed red line in (b) in order to highlight the position of the gates. Note that the amplitudes of the L and R densities are different because the dots have different sizes.}
    \label{fig:bulkasym}
\end{figure*}

We finally illustrate the case of asymmetric quantum dots in the bulk heterostructure. In the following example, we bias differently the side gates of the L and R dots ($V_\mathrm{L}=-24.2$\,mV, $V_\mathrm{R}=-46.9$\,mV, $V_\mathrm{T}=-9.8$\,mV, $V_{\mathrm{S}_2}=V_{\mathrm{S}_4}=40$\,meV). The L and R orbitals of this DQD are plotted in Fig.~\ref{fig:bulkasym} ($\ell_x=20.5$\,nm, $\ell_y=18.7$\,nm in the L dot, and $\ell_x=20.1$\,nm, $\ell_y=13.6$\,nm in the R dot).

The Hamiltonian parameters at this bias point are:
\begin{gather}
    g_\mathrm{L}=
    \begin{pmatrix}
        0.997 & 0     & 1.205 \\
        0     & 0.784 & 0 \\
        1.498 & 0     & 0.651 
    \end{pmatrix},\,
    g_\mathrm{R}=
    \begin{pmatrix}
        0.403 & 0     & -0.515 \\
        0     & 0.641 & 0 \\
       -1.292 & 0     & 1.300 
    \end{pmatrix}; \nonumber \\
    g_\mathrm{T}=
    \begin{pmatrix}
        0.118 & 0 & -0.090 \\
        0 & 0.570 & 0 \\
       -0.032 & 0 & 0.373 
    \end{pmatrix},\,
    \vec{\mu}_\mathrm{T}=
    \begin{pmatrix}
        0 \\
        -17.024 \\
        0 
    \end{pmatrix}\frac{\mu\mathrm{eV}}{\mathrm{T}}; \nonumber \\
    t_c/h=3.5\,\mathrm{GHz},\,\theta_\mathrm{so}=61.55\degree,\,\vec{n}_\mathrm{so}=\vec{y}.
\end{gather}

\begin{figure*}[!h]
    \centering
    \includegraphics[width=\textwidth]{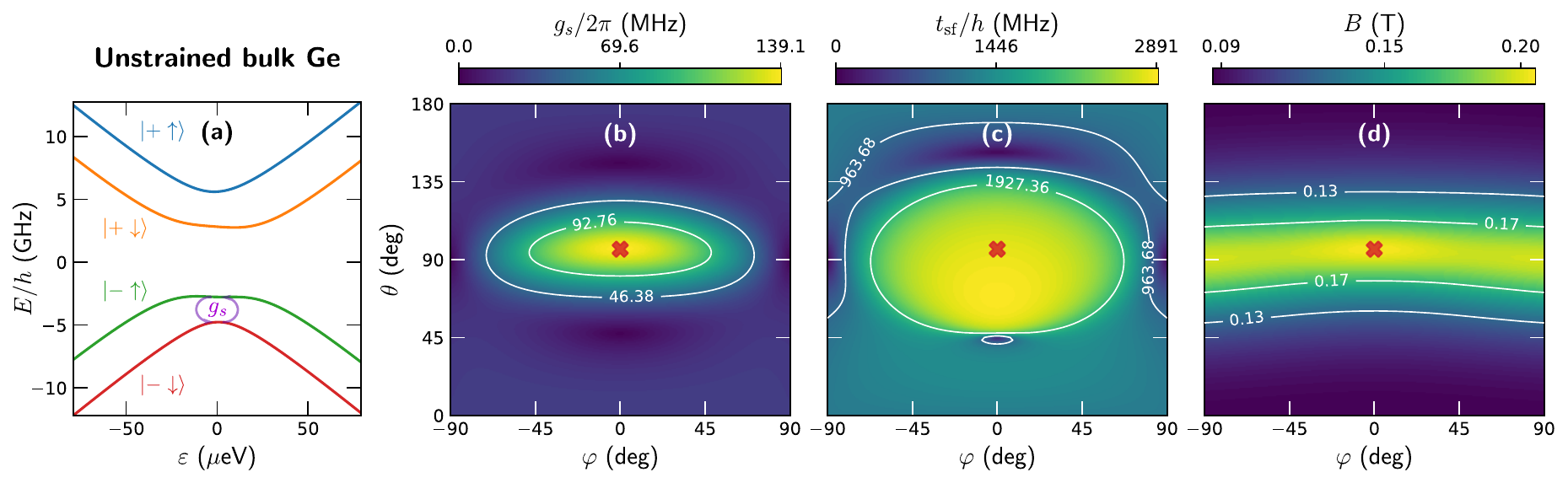}
    \caption{Spin-photon coupling in the bulk Ge heterostructure ($V_\mathrm{L}=-24.2$\,mV, $V_\mathrm{R}=-46.9$\,mV, $V_\mathrm{T}=-9.8$\,mV, $V_{\mathrm{S}_2}=V_{\mathrm{S}_4}=40$\,meV): (a) Spectrum of the DQD as a function of detuning energy $\varepsilon$ ($\vec{B}\parallel\vec{x}$). The resonator couples to the hole in the $\ket{-\uparrow}$ and $\ket{-\downarrow}$ states; (b) Spin-photon coupling $g_s$, (c) spin-flip tunnel coupling $t_\mathrm{sf}$, (d) magnetic field amplitude at resonance ($\omega_s/2\pi=2$\,GHz) as a function of the orientation of $\vec{B}$. The red cross highlights the optimal orientation.}
    \label{fig:gsbulkasym}
\end{figure*}

The maps of spin-photon coupling $g_s$ at zero detuning, the spin-flip tunnel matrix element $t_\mathrm{sf}$ and the magnetic field amplitude at resonance ($\omega_s/2\pi=2$\,GHz) are plotted as a function of the orientation of $\vec{B}$ in Fig.~\ref{fig:gsbulkasym} (along with the hole spectrum as a function of detuning).

The main features of these maps remain similar to Fig.~2e-h of the main text (and Fig.~\ref{fig:gsbulksym}) despite the mismatch between the diagonal elements of $g_\mathrm{L}$ and $g_\mathrm{R}$. The physics indeed remains dominated by the tilt of the gyromagnetic axes (the off-diagonal elements $g_{zx}$ and $g_{xz}$) and by SO tunneling (the angle $\theta_\mathrm{so}$). The optimal $g_s/2\pi=139.14$\,MHz is, however, slightly shifted out-of plane (at $\theta=96\degree$) by this mismatch. Such shifts may vary across different DQDs but shall remain manageable given the width of the features ($g_s/2\pi=129.1$\,MHz being only $7\%$ smaller when $\vec{B}\parallel\vec{x}$). The optimal magnetic field orientation may also slightly move along the $\varphi$ axis if the $(xz)$ symmetry plane is lost.

As expected, the spectrum of the DQD is not symmetric any more with respect to detuning (Fig.~\ref{fig:gsbulkasym}a) as the dots are different. Consequently, the sweet spot in detuning (where $\partial\omega_s/\partial\varepsilon$ is zero) moves away from $\varepsilon=0$. For $\vec{B}\parallel\vec{x}$, there is for example a sweet spot at $\varepsilon=16.5$\,$\mu$eV where the spin-photon coupling at resonance $\omega_s/2\pi=2$\,GHz is $g_s/2\pi=141.34$\,MHz (and is thus slightly larger than at zero detuning). At this sweet spot, the DQD is protected (to first-order) from detuning noise, and, to a large extent from noise on the plunger gates \cite{yu2022strong,rodriguezmena2025}.

\section{Rashba spin-orbit interactions in bulk heterostructures}
\label{sec:SOC}

In this section, we discuss the SO interactions emerging in bulk, unstrained germanium in the two-dimensional (2D) and one-dimensional (1D) limits.

The spectrum of a hole confined in a bulk Ge substrate by a homogeneous vertical electric field $E_z$ can be sorted into subbands $E_n(k_x,k_y)$, where $\vec{k}=(k_x,k_y)$ is the in-plane wave vector and $n$ is a band index. While these subbands are ``spin'' degenerate at $\vec{k}=\vec{0}$ (Kramers degeneracy), they may split at finite $k_x$ and $k_y$ due to SO coupling. The splitting between the lowest two hole subbands can actually be modeled by the following cubic Rashba Hamiltonian acting in the ground HH doublet subspace \cite{Marcellina17,Terrazos21,Rodriguez2023}:
\begin{align}
H_\mathrm{cubic}(\vec{k})&=\beta_R^\prime(k_x^3\sigma_y+k_y^3\sigma_x)+\beta_R(k_yk_xk_y\sigma_y+k_xk_yk_x\sigma_x) \\
& = i\alpha_{R2}(k_-^3\sigma_+-k_+^3\sigma_-)+i\alpha_{R3}(k_-k_+k_-\sigma_--k_+k_-k_+\sigma_+)\,,
\label{eq:H3_cubic}
\end{align}
where $k_\pm=k_x\pm ik_y$ and $\sigma_\pm=\sigma_x\pm i\sigma_y$. The coefficients $\beta_R$ and $\beta_R^\prime$ [or alternatively $\alpha_{R2}=(\beta_R-\beta_R')/4$ and $\alpha_{R3}=(\beta_R+3\beta_R')/4$] characterize the strength of these interactions, which result from the admixture of LH components into the HH doublet by the $R_\mathrm{K}$ and $S_\mathrm{K}$ terms of the LK Hamiltonian [Eq.~\eqref{eq:PQRSkin}].

\begin{figure*}[!h]
    \centering
    \includegraphics[width=0.5\textwidth]{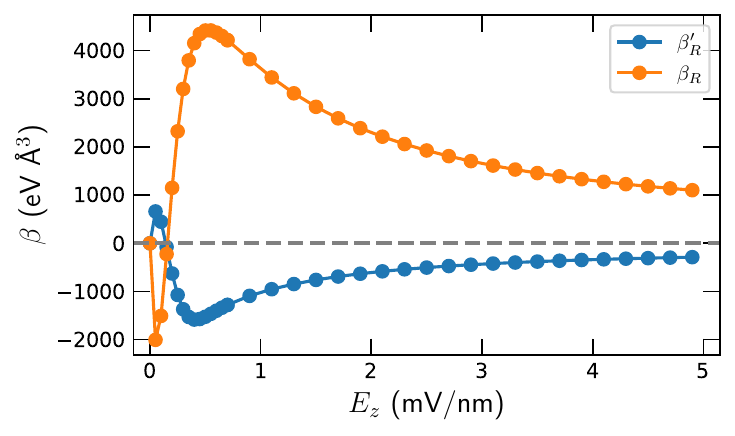}
    \caption{The cubic Rashba coefficients $\beta_R$ and $\beta_R^\prime$ as a function of the vertical electric field $E_z$ in an unstrained, bulk Ge substrate capped with a 20-nm-thick Ge$_{0.8}$Si$_{0.2}$ layer.}
    \label{fig:cubicSO}
\end{figure*}

The coefficients $\beta_R$ and $\beta_R^\prime$ can be extracted from the LK band structure of a bulk Ge/GeSi heterostructure as outlined in Ref.~\cite{Rodriguez2023}. They are plotted\footnote{We practically considered a 100\,nm thick, unstrained Ge film capped with a biaxially strained, 20-nm-thick Ge$_{0.8}$Si$_{0.2}$ layer.} as a function of the vertical electric field $E_z$ in Fig.~\ref{fig:cubicSO}. They can be compared to the coefficients extracted in a strained, 16-nm-thick Ge quantum well by Ref.~\cite{Rodriguez2023}. $\beta_R$ and $\beta_R^\prime$ both tend to zero when $E_z\to 0$ as the inversion symmetry, incompatible with Rashba interactions, is restored. They peak at $E_z\approx 0.5$\,mV/nm, then decrease as the HH/LH bandgap $\Delta$ is opened by the electric field. In the range $1<E_z<5$\,mV/nm, $\beta_R$ is two to three orders of magnitude larger in unstrained, bulk Ge than in strained quantum wells \footnote{Note that $\beta_R$ and $\beta_R^\prime$ have generally opposite signs in this work and in Ref.~\cite{Rodriguez2023}, as we are assuming positive hole dispersion, while Ref.~\cite{Rodriguez2023} assumes negative hole dispersion.}. The Rashba interactions in strained quantum wells are indeed limited by the large HH/LH bandgap $\Delta\approx 70$\,meV opened by biaxial strains, and by the poor response of the structurally confined HH/LH envelopes to the vertical electric field (weak inversion symmetry breaking) \cite{Michal21}. On the opposite, the HH/LH bandgap is exclusively controlled by the electric field in unstrained, bulk heterostructures (it is only $\Delta=4.3$\,meV at $E_z=0.5$\,mV/nm), and the electrical polarizability of the weakly confined HH/LH envelopes is much larger. 

The effects of cubic Rashba interactions on the spin dynamics in quantum dots is dependent on in-plane anisotropies due to partial cancellations between $\beta_R$ and $\beta_R^\prime$ \cite{Marcellina17,Terrazos21}. For disk-shape (circular) quantum dots, they are proportional to $\alpha_{R3}\propto\gamma_2-\gamma_3$ to first order (lattice anisotropy). They can be strongly enhanced by confinement in a squeezed quantum dot with characteristic length scales $\ell_z\ll\ell_y\ll\ell_x$ \cite{Bosco21b}. This indeed gives rise to an effective linear Rashba Hamiltonian $H_\mathrm{linear}=\alpha_R k_x\sigma_y$, with
\begin{equation}
\alpha_R\approx\beta_R\langle k_y^2\rangle 
\label{eq:alphaR}
\end{equation}
and $\langle k_y^2\rangle$ the expectation value of $-\partial^2/\partial y^2$ over the ground-state HH envelope \cite{Marcellina17,Michal21}. We may alternatively introduce the SO length
\begin{equation}
\ell_\mathrm{so}=\frac{\hbar^2}{m^*\alpha_R}\,,
\end{equation}
with $m^*\approx 0.1\,m_0$ the in-plane mass of the hole, and write
\begin{equation}
H_\mathrm{linear}=\frac{\hbar^2}{m^*\ell_\mathrm{so}}k_x\sigma_y\equiv\frac{\hbar}{\ell_\mathrm{so}}v_x\sigma_y\,,
\end{equation}
where $v_x=\hbar k_x/m^*$ is the velocity. If a hole is rigidly translated over a distance $\ell$ along $x$, the corresponding evolution operator
\begin{equation}
U_\mathrm{linear}=e^{-i(\ell/\ell_\mathrm{so})\sigma_y}
\end{equation}
is a spin rotation of angle $2\ell/\ell_\mathrm{so}$ around $y$. This rotation is accounted for by the tunneling angle $\theta_\mathrm{so}\approx s/\ell_\mathrm{so}$ and vector $\vec{n}_\mathrm{so}=\vec{y}$ in the Hamiltonian (1) of the main text, with $s$ the inter-dot distance.

\begin{figure*}[!t]
    \centering
    \includegraphics[width=0.8\textwidth]{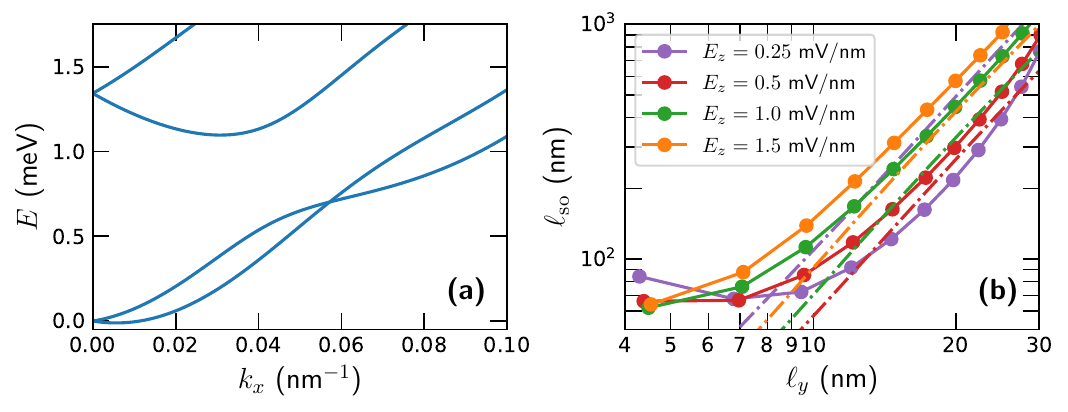}
    \caption{(a) Band structure of a harmonic nanowire in a bulk heterostructure. The hole is confined vertically by a homogeneous electric field $E_z=0.5$\,mV/nm and along $y$ by a harmonic potential with characteristic length scale $\ell_y=\sqrt{\langle y^2\rangle-\langle y\rangle^2}=15$\,nm. The ground-state subbands are split by a linear Rashba interaction. (b) Linear Rashba SO length $\ell_\mathrm{so}$ in such harmonic nanowires as a function of the lateral confinement length $\ell_y$ for different vertical electric fields $E_z$. The dash-dotted lines are the SO lengths expected from the cubic Rashba interaction (Fig.~\ref{fig:cubicSO}) using Eq.~\eqref{eq:alphaR}.}
    \label{fig:linearSO}
\end{figure*}

Eq.~\eqref{eq:alphaR} does not hold, however, if the lateral confinement energy $\hbar^2\langle k_y^2\rangle/(2m^*)$ becomes comparable to the subband splitting \cite{Kloeffel11,Kloeffel18}. This regime can be easily achieved in unstrained heterostructures where the HH/LH bandgap is small. The effective Rashba Hamiltonian $H_\mathrm{linear}$ must then be parametrized directly on the spin splittings in 1D channels. For that purpose, we compute the band structures for parabolic confinement along $y$ with characteristic length scale $\ell_y=\sqrt{\langle y^2\rangle-\langle y\rangle^2}$. The ground-state subband (see Fig.~\ref{fig:linearSO}a) typically exhibits a strong, linear-in-$k$ spin splitting a small $k_x$, and a pattern of higher-order interactions (anti-crossings) at large $k_x$. The extracted $\alpha_R$ (half the slope of the spin splitting at $k_x\to 0$) and SO length $\ell_\mathrm{so}$ are plotted as a function of $\ell_y$ in Fig.~\ref{fig:linearSO}b, for different vertical electric fields $E_z$. As expected, $\alpha_R\propto 1/\ell_\mathrm{so}$ approaches (but remains smaller than) Eq.~\eqref{eq:alphaR} for large $\ell_y\gtrsim 20$\,nm and departs from this trend in narrow channels, except for the smallest electric fields $E_z<0.5$\,mV/nm (where the linear Rashba interactions can, on the opposite, be stronger than expected from the cubic Rashba coefficients up to $\ell_y=30$\,nm). At given $E_z$, $\alpha_R$ shall reach a maximum (and $\ell_\mathrm{so}$ a minimum) for some optimal $\ell_y$ (as lateral confinement ultimately suppresses HH/LH mixing). For $E_z=0.25$\,mV/nm and $\ell_y=14$\,nm representative of the bulk device of Fig.~2 of the main text (namely, achieving similar vertical and lateral extensions as the L/R orbitals), $\ell_\mathrm{so}=114$\,nm is actually comparable to the net distance between the dots (and consistent with the extracted $\theta_\mathrm{so}\approx 60\degree$).

In principle, the uniaxial and shear strains imprinted by the thermal contraction of the metal gates can also give rise to inhomogeneous Rashba- and Dresselhaus-like interactions \cite{Abadillo2023}. However, as discussed in section~\ref{sec:symmetricsqueezed}, the calculated $\theta_\mathrm{so}$ is only weakly dependent on these strains\footnote{More precisely, on the action of the $R_\varepsilon$ and $S_\varepsilon$ terms of Eq.~\eqref{eq:PQRSstrains}.}, at variance with the $g$-matrices $g_\mathrm{L}$ and $g_\mathrm{R}$ (and, to a lesser extent, $g_\mathrm{T}$ and $\vec{\mu}_\mathrm{T}$). We conclude, therefore, that inhomogeneous strains have a much stronger impact on the gyromagnetic response of the single dots than on SO tunneling (partly because the shear strains gradient changes sign from the center of one dot to the other) \cite{MRodriguez2025}.

We emphasize, though, that the 1D limit does not strictly apply to DQDs (even though, as discussed in section~\ref{sec:strainedQW}, they show some fingerprints of quasi-1D dots). It is, however, particularly difficult to de-embed the different SO mechanisms at work in the bulk heterostructures where the electric fields are highly inhomogeneous and the motions along $x$, $y$, and $z$ appear strongly coupled. Moreover, such a break down into generic interactions (Rashba, Dresselhaus, $g$-matrix modulations..) relies on low-order perturbation theory, which hardly applies here given the strong HH/LH mixing. Nevertheless, the above 1D and 2D limits highlight the emergence of strong SO interactions compatible with large tunneling angles $\theta_\mathrm{so}$ between mostly $j_z=\pm3/2$ states.

\section{Effects of a GeSi insertion below the quantum dots}
\label{sec:insertion}

As discussed in sections \ref{sec:parameters} and \ref{sec:SOC}, the SO coupling can be so strong in bulk heterostructures that the typical length scale for spin flips is smaller than the inter-dot distance ($t_\mathrm{sf}$ would thus be larger with smaller SO coupling). The SO coupling strength can be controlled by electrical confinement, by reducing the distance between the dots (which can however make fabrication more difficult), or by heterostructure engineering \cite{delvecchio2026}. 

Here we demonstrate how a thin (5-nm-thick) Ge$_{0.8}$Si$_{0.2}$ insertion below the dots allows for optimal operation. Such an insertion better decouples vertical and lateral confinements, slightly opens the HH/LH bandgap and reduces the SO interactions.

\begin{figure*}[!t]
    \centering
    \includegraphics[width=\textwidth]{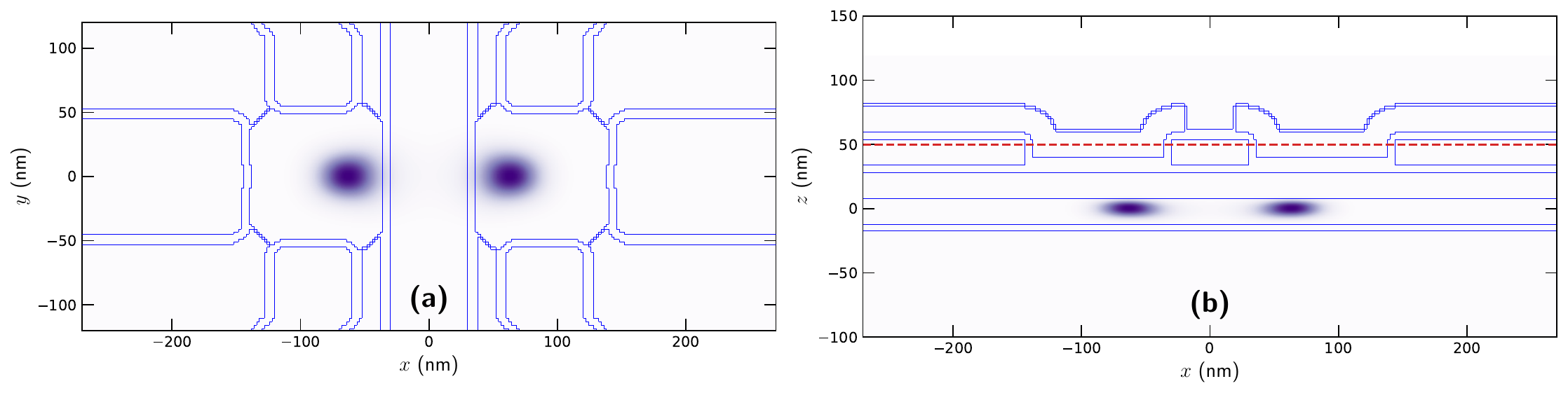}
    \caption{Map of the squared L and R orbitals of the bulk heterostructure with a GeSi insertion in (a) the $(xy)$ plane at $z=0$, and (b) the $(xz)$ plane at $y=0$. The bias point is $V_\mathrm{L}=V_\mathrm{R}=-42.79$\,mV, $V_\mathrm{T}=-12.31$\,mV, $V_{\mathrm{S}_1}=V_{\mathrm{S}_2}=V_{\mathrm{S}_3}=V_{\mathrm{S}_4}=30$\,mV. The blue lines outline the interfaces between the different materials; in panel (a) the materials are plotted in the plane shown by the dashed red line in (b) in order to highlight the position of the gates. The GeSi insertion is visible on panel (b) at coordinate $z\approx-15$\,nm.}
    \label{fig:bulkinsertion}
\end{figure*}

The heterostructure and L and R orbitals of this DQD are plotted in Fig.~\ref{fig:bulkinsertion}. The thin Ge$_{0.8}$Si$_{0.2}$ insertion buried 20\,nm below the top Ge$_{0.8}$Si$_{0.2}$ barrier is well visible on panel (b). The bias point is similar to Fig.~2e-h of the main text. The $g$-matrices and tunneling parameters extracted in the canonical basis set\footnote{The squared norms of the $(\varphi_{+3/2},\varphi_{+1/2},\varphi_{-1/2},\varphi_{-3/2})$ envelopes [Eq.~\eqref{eq:envelopes}] of the L and R $\ket{+\tfrac{3}{2}}$ states are $(96.51\%,1.79\%,1.52\%,0.18\%)$. Those of the time-reversal symmetric $\ket{-\tfrac{3}{2}}$ states are $(0.18\%,1.52\%,1.79\%,96.51\%)$.} are:
\begin{gather}
    g_\mathrm{L}=
    \begin{pmatrix}
        0.413 & 0     & 0.155 \\
        0     & 0.575 & 0 \\
        0.488 & 0     & 4.236 
    \end{pmatrix},\,
    g_\mathrm{R}=
    \begin{pmatrix}
        0.413 & 0     & -0.155 \\
        0     & 0.575 & 0 \\
       -0.488 & 0     & 4.236 
    \end{pmatrix}; \nonumber \\
    g_\mathrm{T}=
    \begin{pmatrix}
        0.037 & 0      & 0 \\
        0     & -0.020 & 0 \\
        0     &        0 & 0.328 
    \end{pmatrix},\,
    \vec{\mu}_\mathrm{T}=
    \begin{pmatrix}
        0 \\
        -2.411 \\
        0 
    \end{pmatrix}\frac{\mu\mathrm{eV}}{\mathrm{T}}; \nonumber \\
    t_c/h=3.5\,\mathrm{GHz},\,\theta_\mathrm{so}=11.60\degree,\,\vec{n}_\mathrm{so}=\vec{y}.
\end{gather}

\begin{figure*}[!t]
    \centering
    \includegraphics[width=\textwidth]{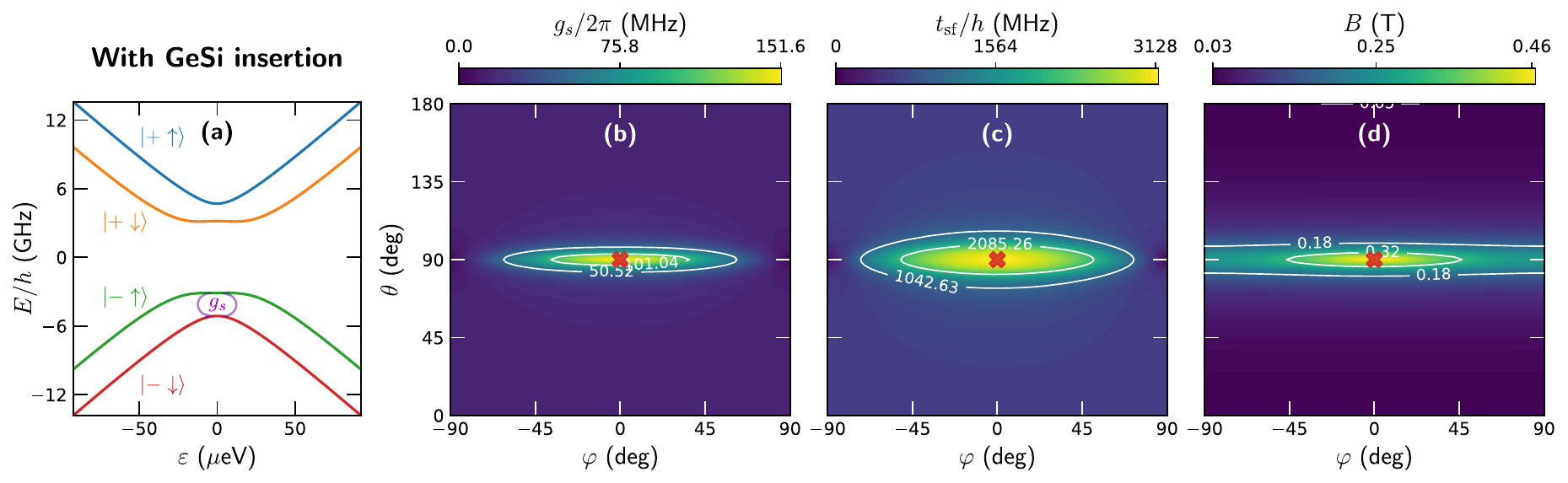}
    \caption{Spin-photon coupling in the bulk Ge heterostructure with a GeSi insertion  ($V_\mathrm{L}=V_\mathrm{R}=-42.79$\,mV, $V_\mathrm{T}=-12.31$\,mV, $V_{\mathrm{S}_1}=V_{\mathrm{S}_2}=V_{\mathrm{S}_3}=V_{\mathrm{S}_4}=30$\,mV): (a) Spectrum of the DQD as a function of detuning energy $\varepsilon$ ($\vec{B}\parallel\vec{x}$). The resonator couples to the hole in the $\ket{-\uparrow}$ and $\ket{-\downarrow}$ states; (b) Spin-photon coupling $g_s$, (c) spin-flip tunnel coupling $t_\mathrm{sf}$, (d) magnetic field amplitude at resonance ($\omega_s/2\pi=2$\,GHz) as a function of the orientation of $\vec{B}$. The red cross highlights the optimal orientation.}
    \label{fig:gsbulkinsertion}
\end{figure*}

The maps of spin-photon coupling $g_s$ at zero detuning, the spin-flip tunnel matrix element $t_\mathrm{sf}$ and the magnetic field amplitude at resonance ($\omega_s/2\pi=2$\,GHz) are plotted as a function of the orientation of $\vec{B}$ in Fig.~\ref{fig:gsbulkinsertion}.

The spin-photon coupling reaches $g_s/2\pi=151.57$\,MHz for $\vec{B}\parallel\vec{x}$. The LH mixing is only 3.12\%, which softens the effects of SO coupling. As a result of the stronger vertical confinement, the motions along $x$, $y$ and $z$ are better decoupled, which further reduces $g_{xz}$ and $g_{zx}$ with respect to the plain device without insertion [Eqs.~\eqref{eq:bulkparams}]. The rotations of the principal magnetic and spin axes of each dot are thus only $\delta\theta_\mathrm{R}^m=-\delta\theta_\mathrm{L}^m=6.82\degree$ and $\delta\theta_\mathrm{R}^s=-\delta\theta_\mathrm{L}^s=0.05\degree$. On the other hand, the gyromagnetic anisotropy is larger, as expected for more heavy-hole-like states (the principal $g$-factors of $g_\mathrm{L}$ and $g_\mathrm{R}$ are $g_1=0.392$, $g_2=0.574$, $g_3=4.269$). This mildly shrinks all features on the maps of Fig.~\ref{fig:gsbulkinsertion}; the peak of $g_s$ is, nevertheless, still reasonably broad (FWHM $\Delta\theta=9\degree$ at $\varphi=0$). 

The tunneling angle $\theta_\mathrm{so}=11.6\degree$ is also reduced by the GeSi insertion (the Rashba interactions being weaker due to the increase of the HH/LH bandgap and decrease of the electrical polarizability of the envelopes). As a consequence, the angle between the Larmor vectors $\vec{\omega}_\mathrm{L}=\mu_B g_\mathrm{L}\vec{x}/\hbar$ and $\vec{\omega}_\mathrm{R}=\mu_B g_\mathrm{R}\vec{x}/\hbar$ is $\Theta_\mathrm{LR}=99.47\degree$, while $\widehat{\Theta}_\mathrm{LR}=\Theta_\mathrm{LR}+2\theta_\mathrm{so}=122.67\degree$. Therefore, the spin-flip tunnel matrix element $t_\mathrm{sf}=t_c\mathrm{sin}(\widehat{\Theta}_\mathrm{LR}/2)=0.88t_c$ ($\vec{B}\parallel\vec{x}$) is larger than in the plain Ge substrate ($t_\mathrm{sf}=0.77t_c$ at comparable bias). 

We emphasize, though, that trying to reach even larger $t_\mathrm{sf}/t_c$ ultimately comes with a strong decrease of the net $g$-factors of the DQD (as the $\ket{-\sigma}$ states are balanced mixtures of $\ket{\Uparrow}$ components in one dot and $\ket{\Downarrow}$ components in the other), thus with a strong increase of the magnetic field needed to reach the desired $\omega_s$ (see demonstration in section \ref{sec:frames}). It appears safer, therefore, to target $t_\mathrm{sf}\approx 0.9t_c$ -- a limit reached with the present insertion. This demonstrates that heterostructure engineering can, if necessary, be used to further optimize bulk germanium devices.

\section{Frames and general expressions for spin-photon coupling}
\label{sec:frames}

In this section, we discuss the different frames in more details, and provide analytical expressions for the spin-photon coupling, which give further insights into the trends highlighted in this work.

At zero magnetic field, the ground level of each isolated dot is a twofold-degenerate Kramers doublet. Mapping these doublets onto the pseudo-spins $\Uparrow,\,\Downarrow$ of Eq.~(1) of the main text therefore involves an arbitrary choice: any change of local pseudo-spin basis
\begin{equation}
    U=U_{\rm L}\,\tau_{\rm L}+U_{\rm R}\,\tau_{\rm R}\,,\qquad U_{\rm L},\,U_{\rm R}\in\mathrm{SU}(2)\,,
    \label{eq:smgauge}
\end{equation}
yields an equally valid four-state model, with parameters transformed as
\begin{equation}
    g_\mathrm{L}\to\hat{g}_\mathrm{L}=\mathcal{R}_\mathrm{L}(\theta_\mathrm{L},\vec{n}_\mathrm{L})g_\mathrm{L}\,,\qquad g_\mathrm{R}\to\hat{g}_\mathrm{R}={\cal R}_\mathrm{R}(\theta_\mathrm{R},\vec{n}_\mathrm{R})g_\mathrm{R}\,,\qquad
    T\to U_{\rm L}^\dagger\,T\,U_{\rm R}\,,
    \label{eq:smgaugetransf}
\end{equation}
where $\mathcal{R}_i(\theta_\mathrm{i},\vec{n}_\mathrm{i})\in\mathrm{SO}(3)$ is the rotation of angle $\theta_\mathrm{i}$ around $\vec{n}_\mathrm{i}$ associated with $U_i$ [$U_i^\dagger\vec{\sigma}U_i=\mathcal{R}_i\vec{\sigma}$], and $T=-t_c\,e^{-i\theta_\mathrm{so}\vec{n}_\mathrm{so}\cdot\vec{\sigma}}+\frac{\mu_B}{2}\,\vec{\sigma}\cdot g_\mathrm{T}\vec{B}-\frac{i}{2}(\boldsymbol{\mu}_{\rm T}\cdot\vec{B})\,\sigma_0$ collects the tunneling and magnetotunneling terms [the tunneling block of Eq.~(1) being $\tau_+T+\mathrm{h.c.}$, with $\tau_+=\ket{\rm L}\bra{\rm R}$]. All parameters of Eq.~(1) are thus gauge-dependent~\cite{Venitucci18}: the $g$-matrices are defined up to a rotation in each dot, and the decomposition of $T$ into $\theta_\mathrm{so}$, $\vec{n}_\mathrm{so}$, $g_\mathrm{T}$ and $\boldsymbol{\mu}_{\rm T}$ changes from gauge to gauge. The physical content lies in the invariants: the spectrum, the charge tunneling $t_c$ [with $t_c^2=\det T$ at $\vec{B}=\vec{0}$], the Zeeman splittings $\mu_B|g_{\rm L/R}\vec{B}|$, the spin-conserving and spin-flip tunnel amplitudes $t_\mathrm{sc}$ and $t_\mathrm{sf}$ between the Zeeman eigenstates of the uncoupled dots, and the spin-photon couplings defined below. In particular, the SO tunneling angle $\theta_\mathrm{so}$ can be gauged away entirely (see section~\ref{sec:soframe}); the SO physics it encodes then resurfaces in the transformed $g$-matrices. A ``frame'' in this section is nothing but a specific choice of gauge in Eq.~\eqref{eq:smgauge}: moving between frames does not change the physics, but each frame makes a different aspect of the problem transparent.

In particular, the coupling to the resonator threads through all frames unchanged. The photon enters Eq.~(1) through $H_{\rm c}=eV_\mathrm{zpf}(a+a^\dagger)D_{\rm L}$ with $D_{\rm L}\equiv\alpha_{\rm L}\tau_{\rm L}$ acting on the charge sector only; since $\tau_{\rm L}$ commutes with every transformation~\eqref{eq:smgauge}, the coupling operator -- and thus the spin-photon couplings $g_s$ and $g_\parallel$ of Eq.~(2) of the main text, which are matrix elements of $D_{\rm L}$ between eigenstates of the DQD -- can be evaluated in whichever frame is most convenient.  The canonical frame fixes the reference gauge from the microscopic wave functions themselves: the pseudo-spins are chosen as heavy-hole-like $\pm\tfrac{3}{2}$ doublets, so that the fitted $g_{\rm L/R}$, $\theta_\mathrm{so}$, $\vec{n}_\mathrm{so}$, $g_\mathrm{T}$ and $\boldsymbol{\mu}_{\rm T}$ can be analyzed within the framework of perturbation theory for the effective SO interactions in planar Ge (Rashba, Dresselhaus, $g$-TMR~\cite{Venitucci18}); this is the gauge in which all parameters of the main text and supplementary material are quoted. Starting from the canonical frame, we consider two frames with  physical significance for the spin-photon interaction. The spin-orbit frame (section~\ref{sec:soframe}) applies the local rotations that block-diagonalizes the tunneling Hamiltonian at zero magnetic field~\cite{Sen23,Geyer24}, providing an interpretation of the spin-photon coupling in terms of geometrical relationships between Larmor vectors. The local Zeeman frame (section~\ref{sec:zeemanframe}) diagonalizes the Zeeman Hamiltonians of the uncoupled dots at a given magnetic-field orientation, thus providing an insightful connection between the spin-photon coupling and the spin-flip tunneling $t_\mathrm{sf}$ for a given magnetic field orientation~\cite{yu2022strong}.

Both subsections below start from the canonical frame model, Eq.~(1) of the main text, reproduced here for convenience:
\begin{align}
    H&=H_{\rm S}+H_{\rm T}+H_{\rm MT}+H_{\rm r}\,,\nonumber\\
    H_{\rm S}&=\frac{\varepsilon}{2}\tau_z+\frac{\mu_B}{2}\left[(\vec{\sigma}\cdot g_{\rm L}\vec{B})\,\tau_{\rm L}+(\vec{\sigma}\cdot g_{\rm R}\vec{B})\,\tau_{\rm R}\right],\nonumber\\
    H_{\rm T}&=-(t_c\cos\theta_{\rm so})\,\tau_x-(t_c\sin\theta_{\rm so}\,\vec{n}_{\rm so}\cdot\vec{\sigma})\,\tau_y\,,\nonumber\\
    H_{\rm MT}&=\frac{\mu_B}{2}(\vec{\sigma}\cdot g_\mathrm{T}\vec{B})\,\tau_x+\frac{1}{2}(\boldsymbol{\mu}_{\rm T}\cdot\vec{B})\,\tau_y\,,\nonumber\\
    H_{\rm r}&=\hbar\omega_{\rm r}\,a^\dagger a+eV_\mathrm{zpf}\,(a+a^\dagger)\,D_{\rm L}\,,
    \label{eq:smH}
\end{align}
where $\tau_{\rm L,R}=\tfrac{1}{2}(\tau_0\pm\tau_z)$ and $\vec{\omega}_{\rm L/R/T}=\mu_Bg_{\rm L/R/T}\vec{B}/\hbar$ are the Larmor vectors (with angular frequencies $\omega_{\rm L/R}$). Since the $\propto\tau_0$ component of $D_{\rm L}$ merely displaces the photon field, the photon effectively couples to the charge dipole, $H_{\rm c}=\hbar g_{\rm c}(a+a^\dagger)\tau_z+\mathrm{const}$, with $\hbar g_{\rm c}\equiv e\alpha_{\rm L}V_\mathrm{zpf}/2$ the charge-photon coupling entering both calculations below.

\subsection{Spin-orbit frame}
\label{sec:soframe}
The spin-orbit frame gauges the SO tunneling away entirely, transferring it to dressed $g$-matrices. Applying the unitary $U_\mathrm{so}=\exp\left[-i\theta_\mathrm{so}\tau_z\,\vec{n}_\mathrm{so}\cdot\vec{\sigma}/2\right]$ to Eq.~\eqref{eq:smH} [i.e., $U_{\rm L/R}=\exp(\mp i\theta_\mathrm{so}\vec{n}_\mathrm{so}\cdot\vec{\sigma}/2)$ in Eq.~\eqref{eq:smgauge}], we obtain $H^\mathrm{(so)}=U_\mathrm{so}^\dagger HU_\mathrm{so}$ with
\begin{align}
    H^\mathrm{(so)}&=H_{\rm S}^\mathrm{(so)}+H_{\rm T}^\mathrm{(so)}+H_{\rm MT}^\mathrm{(so)}+H_{\rm r}\,,\nonumber\\
    H_{\rm S}^\mathrm{(so)}&=\frac{\varepsilon}{2}\tau_z+\frac{\mu_B}{2}\left[(\vec{\sigma}\cdot \hat{g}_{\rm L}\vec{B})\tau_{\rm L}+(\vec{\sigma}\cdot \hat{g}_{\rm R}\vec{B})\tau_{\rm R}\right],\nonumber\\
    H_{\rm T}^\mathrm{(so)}&=-t_c\tau_x\,,\nonumber\\
    H_{\rm MT}^\mathrm{(so)}&=\frac{\mu_B}{2}(\vec{\sigma}\cdot \hat{g}_{\rm T}\vec{B})\tau_x+\frac{1}{2}(\hat{\boldsymbol{\mu}}_{\rm T}\cdot\vec{B})\,\tau_y\,,
    \label{eq:smHso}
\end{align}
where the local $g$-matrices are rotated in opposite directions, $g_{\rm L/R}\to\hat{g}_{\rm L/R}={\cal R}(\mp\theta_\mathrm{so},\vec{n}_\mathrm{so})g_{\rm L/R}$, and the magnetotunneling block transforms as $\hat{g}_\mathrm{T}=\left[\mathds{1}-(1-\cos\theta_\mathrm{so})\,\vec{n}_\mathrm{so}\otimes \vec{n}_\mathrm{so}\right]g_\mathrm{T}+\sin\theta_\mathrm{so}\,\vec{n}_\mathrm{so}\otimes\boldsymbol{\mu}_\mathrm{T}/\mu_B$ and $\hat{\boldsymbol{\mu}}_\mathrm{T}=\cos\theta_\mathrm{so}\,\boldsymbol{\mu}_\mathrm{T}-\mu_B\sin\theta_\mathrm{so}\,({}^tg_\mathrm{T}\,\vec{n}_\text{so})$~\cite{rodriguezmena2025}.

To estimate the spin-photon coupling in the lowest charge branch $\ket{-}$, we first diagonalize the charge sector at $\vec{B}=\vec{0}$, which is trivial in the SO frame. Introducing the charge qubit energy $\Omega=\sqrt{\varepsilon^2+4t_c^2}$ and the charge mixing angle $\Theta$ [$\cos\Theta=\varepsilon/\Omega$, $\sin\Theta=2t_c/\Omega$], a rotation of the charge sector by $\Theta$ about $\tau_y$ maps the tunneling onto $(\Omega/2)\tau_z$ and yields
\begin{equation}
\begin{aligned}
    H^{\text{(so)}}_{\rm cd}&=\frac{\Omega}{2}\tau_z+\frac{\hbar}{2}\hat{\vec{\omega}}_+\cdot\vec{\sigma}
    +\frac{\hbar}{2}(\cos\Theta\tau_z+\sin\Theta\tau_x)\hat{\vec{\omega}}_-\cdot\vec{\sigma}
    +\frac{\hbar}{2}(\cos\Theta\,\tau_x-\sin\Theta\tau_z)\hat{\vec{\omega}}_{\rm T}\cdot\vec{\sigma}
    +\frac{\hbar}{2}\hat{\nu}_\mathrm{T}\tau_y\\
    &\quad+\hbar\omega_{\rm r} a^\dagger a+\hbar g_{\rm c}(a+a^\dagger)(\cos\Theta\tau_z+\sin\Theta\tau_x)\,,
\end{aligned}
    \label{eq:socharge}
\end{equation}
where we have defined $\hat{\vec{\omega}}_{\pm}=(\hat{\vec{\omega}}_\text{L}\pm\hat{\vec{\omega}}_\text{R})/2$, with $\hat{\vec{\omega}}_{\rm L/R/T}=\mu_B\hat{g}_{\rm L/R/T}\vec{B}/\hbar$ and $\hat{\nu}_\mathrm{T}=\hat{\boldsymbol{\mu}}_{\rm T}\cdot\vec{B}/\hbar$. The lower branch is $\langle\tau_z\rangle=-1$.

This frame is particularly convenient because the charge diagonalization is spin-independent, while both the linear-in-$\vec{B}$ terms and the charge-photon coupling now connect the two branches through their $\tau_x$ components. A Schrieffer-Wolff elimination of the upper branch~\cite{schrieffer1966,bravyi2011}, following Ref.~\cite{rodriguezmena2025}, then yields the effective Hamiltonian of the lower branch. Ignoring spin-independent and purely photonic contributions (displacement and squeezing), it reads
\begin{equation}
  H_{-}^{\rm (so)}=\hbar\tilde{\omega}_{\rm r}a^\dagger a-\hbar g_{\rm c}\cos\Theta(a+a^\dagger) +\frac{\hbar}{2}\left(\hat{\vec{\omega}}_s+\hat{\vec{\omega}}_\text{ad}a^\dagger a\right)\cdot\vec{\sigma}+\hbar\,(\hat{\vec{g}}_s\cdot\vec{\sigma})(a+a^\dagger)\,,
\label{eq:lowso}
\end{equation}
with $\tilde{\omega}_{\rm r}=\omega_{\rm r}-2\hbar g_{\rm c}^2\sin^2\Theta\,\Omega/[\Omega^2-(\hbar\omega_{\rm r})^2]$ the renormalized resonator frequency, $\hat{\vec{\omega}}_s$ the Larmor vector of the lower branch, $\hat{\vec{g}}_s$ the spin-photon coupling vector, and $\hat{\vec{\omega}}_\text{ad}$ the adiabatic curvature correction to the dispersive shift~\cite{park2020adiabatic, chessari2025unifying}. Their explicit expressions are
\begin{align}
    \hat{\vec{\omega}}_s&=\hat{\vec{\omega}}_{+}-\cos\Theta\hat{\vec{\omega}}_{-}+\sin\Theta\hat{\vec{\omega}}_{\rm T}
    +\frac{2g_{\rm c}^2\sin^2\Theta}{(\Omega/\hbar+\omega_{\rm r})^2}\left(\cos\Theta\hat{\vec{\omega}}_{-}-\sin\Theta\,\hat{\vec{\omega}}_{\rm T}\right) \nonumber \\
    &\quad+\frac{2g_{\rm c}^2\sin\Theta\cos\Theta\,(\hbar\omega_{\rm r}-2\Omega)}{\Omega\left[\Omega^2/\hbar^2-\omega_{\rm r}^2\right]}\left(\sin\Theta\,\hat{\vec{\omega}}_{-}+\cos\Theta\,\hat{\vec{\omega}}_{\rm T}\right),
    \label{eq:smws}\\
    \hat{\vec{g}}_s&=-\frac{\hbar g_{\rm c}}{2}\left[\frac{1}{\Omega}+\frac{\Omega}{\Omega^2-(\hbar\omega_{\rm r})^2}\right]\sin\Theta\left(\sin\Theta\hat{\vec{\omega}}_{-}+\cos\Theta\hat{\vec{\omega}}_{\rm T}\right),
    \label{eq:smgsvec}\\
    \hat{\vec{\omega}}_\text{ad}&=2(\hbar g_{\rm c})^2\sin^2\Theta\left[
    \frac{1}{(\Omega-\hbar\omega_{\rm r})^2}+\frac{1}{(\Omega+\hbar\omega_{\rm r})^2}\right]
    \left(\cos\Theta\hat{\vec{\omega}}_{-}-\sin\Theta\hat{\vec{\omega}}_{\rm T}\right) & \nonumber \\
    &\quad+\frac{8(\hbar g_{\rm c})^2\sin\Theta\cos\Theta}{\Omega^2-(\hbar\omega_{\rm r})^2}
    \left(\sin\Theta\hat{\vec{\omega}}_{-}+\cos\Theta\hat{\vec{\omega}}_{\rm T}\right)\,.
    \label{eq:smwad}
\end{align}
Note that the spin-photon coupling is a vector at this stage. To separate it into the longitudinal ($g_\parallel$) and transverse ($g_s$) interactions of Eq.~(2) of the main text, $\hat{\vec{g}}_s$ needs to be projected onto and perpendicular to the Larmor vector of the lower branch:
\begin{equation}
    g_\parallel=\frac{\hat{\vec{\omega}}_s\cdot \hat{\vec{g}}_s}{\hat{\omega}_s}\,,\qquad
    g_s=\frac{|\hat{\vec{\omega}}_s\times \hat{\vec{g}}_s|}{\hat{\omega}_s}\,,
    \label{eq:smproj}
\end{equation}
with $\hat{\omega}_s=|\hat{\vec{\omega}}_s|$ the spin frequency (simply noted $\omega_s$ in the main text).

Some interesting limits can be explored. At zero detuning ($\cos\Theta=0$, $\sin\Theta=1$, $\Omega=2t_c$), the qubit axis $\hat{\vec{\omega}}_s\approx\hat{\vec{\omega}}_++\hat{\vec{\omega}}_{\rm T}$ is set by the mean dressed Larmor vector with a magnetotunneling tilt, whereas the coupling vector $\hat{\vec{g}}_s=-(\hbar g_{\rm c}\Lambda_0/2)\,\hat{\vec{\omega}}_-$ [with $\Lambda_0=(2t_c)^{-1}+2t_c/[4t_c^2-(\hbar\omega_{\rm r})^2]$] is proportional to the half-difference $\hat{\vec{\omega}}_-$ only: magnetotunneling drops out of the coupling operator at $\varepsilon=0$, but may still rotate the axis onto which it is projected. The spin-photon interactions are, in this case,
\begin{equation}
    g_\parallel=-\frac{\hbar g_{\rm c}\Lambda_0}{2}\,\frac{(\hat{\vec{\omega}}_++\hat{\vec{\omega}}_{\rm T})\cdot\hat{\vec{\omega}}_-}{|\hat{\vec{\omega}}_++\hat{\vec{\omega}}_{\rm T}|}\,,\qquad
    g_s=\frac{\hbar g_{\rm c}\Lambda_0}{2}\,\frac{|(\hat{\vec{\omega}}_++\hat{\vec{\omega}}_{\rm T})\times\hat{\vec{\omega}}_-|}{|\hat{\vec{\omega}}_++\hat{\vec{\omega}}_{\rm T}|}\,.
    \label{eq:smso0}
\end{equation}
In the limit $\hat{\vec{\omega}}_{\rm T}=\vec{0}$, these can be connected to the relative angle $\widehat{\Theta}_\mathrm{LR}$ between the dressed Larmor vectors, with $\cos\widehat{\Theta}_\mathrm{LR}=\hat{\vec{\omega}}_{\rm L}\cdot\hat{\vec{\omega}}_{\rm R}/(\omega_{\rm L}\omega_{\rm R})$. Using $\hat{\vec{\omega}}_+\cdot\hat{\vec{\omega}}_-=(\omega_{\rm L}^2-\omega_{\rm R}^2)/4$ and $|\hat{\vec{\omega}}_+\times\hat{\vec{\omega}}_-|=(\omega_{\rm L}\omega_{\rm R}\sin\widehat{\Theta}_\mathrm{LR})/2$,
\begin{equation}
    g_\parallel=-\frac{\hbar g_{\rm c}\Lambda_0}{8}\,\frac{\omega_{\rm L}^2-\omega_{\rm R}^2}{\omega_s}\,,\qquad
    g_s=\frac{\hbar g_{\rm c}\Lambda_0}{4}\frac{\omega_{\rm L}\omega_{\rm R}\sin\widehat{\Theta}_\mathrm{LR}}{\omega_s}\,.
    \label{eq:smso0ang}
\end{equation}
For quasi-symmetric dots ($\omega_{\rm L}\approx\omega_{\rm R}\equiv\bar{\omega}$), $g_\parallel$ vanishes, $\omega_s=\bar{\omega}{\rm cos}(\widehat{\Theta}_\mathrm{LR}/2)$, and $g_s=(\hbar g_{\rm c}\Lambda_0/2)\,\bar{\omega}{\rm sin}(\widehat{\Theta}_\mathrm{LR}/2)=(\hbar g_{\rm c}\Lambda_0/2)\,\bar{\omega}\,t_\mathrm{sf}/t_c$ by Eq.~\eqref{eq:smtsctsf}: in the adiabatic charge limit $\hbar\omega_{\rm r}\ll2t_c$ ($\Lambda_0\to1/t_c$), this reduces to $g_s\approx e\alpha_{\rm L}V_\mathrm{zpf}(\omega_{\rm L}+\omega_{\rm R})t_\mathrm{sf}/(8t_c^2)$, which is Eq.~(3) of the main text. 

We emphasize that Eqs.~\eqref{eq:smso0} and \eqref{eq:smso0ang} are proportional to the applied magnetic field (as expected since time-reversal symmetry must be broken to couple a spin to an electric field). In the devices of Fig.~2 of the main text, $\hat{\vec{\omega}}_{\rm T}\ne\vec{0}$ but Eq.~\eqref{eq:smso0ang} still holds for $\vec{B}\parallel\vec{x}$ as $\hat{\vec{\omega}}_{\rm T}\parallel\hat{\vec{\omega}}_+$. Yet $\hat{\vec{\omega}}_{\rm T}$ also enters in $\omega_s$ [Eq.~\eqref{eq:smws}]: the magnetic field needed to reach a target spin frequency is modulated by magnetotunneling, and so is $g_s$ {\it at constant $\omega_s$}. This explains why magnetotunneling corrections do enhance $g_s$ by $\approx 20\%$ on Fig.~(2)f of the main text: at $\omega_s/2\pi=2$\,GHz, the magnetic field is indeed larger with than without magnetotunneling.

\subsection{Local-Zeeman frame}
\label{sec:zeemanframe}
Starting from Eq.~\eqref{eq:smH}, we can rotate the pseudo-spin basis of each individual dot to diagonalize its Zeeman Hamiltonian: we choose $U_{\rm L},U_{\rm R}$ in Eq.~\eqref{eq:smgauge} such that $U_{\rm L/R}^\dagger(\vec{\omega}_{\rm L/R}\cdot\vec{\sigma})U_{\rm L/R}=\omega_{\rm L/R}\sigma_z$. The rotated basis states $\ket*{{\rm L}\tilde{\Uparrow}}$, $\ket*{{\rm L}\tilde{\Downarrow}}$, $\ket*{{\rm R}\tilde{\Uparrow}}$, $\ket*{{\rm R}\tilde{\Downarrow}}$ are the Zeeman eigenstates of the uncoupled dots. The diagonalization fixes $U_{\rm L/R}$ only up to residual rotations about $z$; the latter can always be chosen so that the tunneling Hamiltonian becomes
\begin{equation}
    H_\mathrm{T}=-(t_c\cos\theta_{\rm so})\tau_x-(t_c\sin\theta_{\rm so}\vec{n}_{\rm so}\cdot\vec{\sigma})\tau_y\;\to\;H_\mathrm{T}^{\rm(LZ)}=-t_\mathrm{sc}\tau_x-t_\mathrm{sf}\tau_y\sigma_y\,,
    \label{eq:smHTLZ}
\end{equation}
where $t_\mathrm{sc}$ and $t_\mathrm{sf}$ are the spin-conserving and spin-flip tunnel couplings for the given magnetic field orientation ($t_\mathrm{sf}=\bra*{{\rm L}\tilde{\Uparrow}}H_\mathrm{T}\ket*{{\rm R}\tilde{\Downarrow}}=-\bra*{{\rm L}\tilde{\Downarrow}}H_\mathrm{T}\ket*{{\rm R}\tilde{\Uparrow}}$). They are both real and satisfy $t_\mathrm{sc}^2+t_\mathrm{sf}^2=t_c^2$. This invariant allows for the parameterization
\begin{equation}
    t_\mathrm{sc}=t_c\cos\left(\frac{\widehat{\Theta}_{\rm LR}}{2}\right)\,,\qquad
    t_\mathrm{sf}=t_c\sin\left(\frac{\widehat{\Theta}_{\rm LR}}{2}\right)\,.
    \label{eq:smtsctsf}
\end{equation}
After some algebra it can be shown that $\widehat{\Theta}_{\rm LR}$ is actually the same angle as in section~\ref{sec:soframe} (the angle between the dressed Larmor vectors $\hat{\vec{\omega}}_{\rm L}$ and $\hat{\vec{\omega}}_{\rm R}$), hence the notation. $\widehat{\Theta}_{\rm LR}$ can also be related to the canonical frame parameters; for that purpose we define the unit Larmor vectors $\vec{n}_{\rm L/R}=\vec{\omega}_{\rm L/R}/\omega_{\rm L/R}$, and find:
\begin{equation}
    \cos\widehat{\Theta}_\mathrm{LR}=\cos2\theta_\mathrm{so}(\vec{n}_{\rm L}\cdot\vec{n}_{\rm R})+\sin2\theta_\mathrm{so}\vec{n}_\mathrm{so}\cdot(\vec{n}_{\rm R}\times\vec{n}_{\rm L})+(1-\cos2\theta_\mathrm{so})(\vec{n}_\mathrm{so}\cdot\vec{n}_{\rm L})(\vec{n}_\mathrm{so}\cdot\vec{n}_{\rm R})\,,
    \label{eq:smThetaLR}
\end{equation}
which embeds on the same footing the two sources of spin-flip tunneling: the non-collinearity of the local Larmor vectors and the spin rotation imparted by the SO tunneling [entering with its full angle $2\theta_{\rm so}$, cf.\ Eq.~\eqref{eq:smH}]. In particular, when the SO tunneling can be neglected ($\theta_\mathrm{so}=0$), we recover $\cos\widehat{\Theta}_\mathrm{LR}=\vec{n}_{\rm L}\cdot\vec{n}_{\rm R}$: $\widehat{\Theta}_{\rm LR}$ is then the angle between the bare Larmor vectors, and the spin-flip tunneling entirely results from the imbalance between the $g$-matrices. Conversely, for identical dots with $\vec{B}\perp\vec{n}_\mathrm{so}$ ($\vec{n}_{\rm L}=\vec{n}_{\rm R}\perp\vec{n}_\mathrm{so}$), $\widehat{\Theta}_{\rm LR}=2\theta_\mathrm{so}$ and $t_\mathrm{sf}=t_c\sin\theta_\mathrm{so}$, as can be read directly off Eq.~\eqref{eq:smH}. Finally, if both Larmor vectors are parallel to $\vec{n}_\mathrm{so}$, $t_\mathrm{sf}=0$, as expected: precession axes aligned with the SO axis are unaffected by the tunneling spin rotation.

Moreover, the magnetotunneling Hamiltonian keeps its canonical structure,
\begin{equation}
    H_\text{MT}^{\rm(LZ)}=\frac{\mu_B}{2}\left(\vec{\sigma}\cdot g_\mathrm{T}^{\rm LZ}\vec{B}\right)\tau_x+\frac{1}{2}\left(\boldsymbol{\mu}_{\rm T}^{\rm LZ}\cdot\vec{B}\right)\tau_y\,,
    \label{eq:smHMTLZ}
\end{equation}
with effective responses $g_\mathrm{T}^{\rm LZ}$ and $\boldsymbol{\mu}_{\rm T}^{\rm LZ}$ that now depend on the orientation of the magnetic field through $U_{\rm L/R}$ [whose residual phases are fixed by Eq.~\eqref{eq:smHTLZ}]. Their expressions in terms of the canonical frame $g_\mathrm{T}$ and $\boldsymbol{\mu}_{\rm T}$ are rather long, cumbersome and offer little additional insight, particularly owing to the present choice of gauge in which the spin-flip tunneling term is proportional to $\sigma_y$; we therefore do not present them explicitly.

Collecting all terms, the local-Zeeman equivalent of Eq.~\eqref{eq:smH} reads
\begin{align}
    H^{\rm(LZ)}&=\frac{\varepsilon}{2}\tau_z+\frac{\hbar\bar{\omega}}{2}\sigma_z+\frac{\hbar\delta\omega}{2}\tau_z\sigma_z-t_\mathrm{sc}\tau_x-t_\mathrm{sf}\tau_y\sigma_y\nonumber\\
    &\quad+\frac{\mu_B}{2}\left(\vec{\sigma}\cdot g_\mathrm{T}^{\rm LZ}\vec{B}\right)\tau_x+\frac{1}{2}\left(\boldsymbol{\mu}_{\rm T}^{\rm LZ}\cdot\vec{B}\right)\tau_y+\hbar\omega_{\rm r}a^\dagger a+\hbar g_{\rm c}(a+a^\dagger)\tau_z\,,
    \label{eq:smHLZ}
\end{align}
with $\bar{\omega}=(\omega_{\rm L}+\omega_{\rm R})/2$ and $\delta\omega=(\omega_{\rm L}-\omega_{\rm R})/2$ the mean and half-difference Larmor frequencies.

To estimate the spin-photon coupling in the lowest charge branch using the local-Zeeman parameters, first note that Eqs.~\eqref{eq:smHLZ} and~\eqref{eq:smHso} are related by a transformation of the form~\eqref{eq:smgauge} acting on the spin sector only: $\varepsilon$, $t_c$ and the charge-photon coupling -- hence $\Omega$, $\Theta$ and $\Lambda$ of section~\ref{sec:soframe} -- are common to both frames. The charge sector, however, can no longer be diagonalized on its own: the spin-flip tunneling entangles charge and spin. Rather than repeating the Schrieffer--Wolff construction, we gauge away $t_\mathrm{sf}$ (switch to the SO frame): the effective Hamiltonian of the lower branch is then exactly Eq.~\eqref{eq:lowso}, with the vectors of Eqs.~\eqref{eq:smws}-\eqref{eq:smwad} re-expressed as a function local-Zeeman parameters. In the effective spin basis -- $\vec{e}_z$ along the bisector of $\hat{\vec{\omega}}_{\rm L}$ and $\hat{\vec{\omega}}_{\rm R}$, $\vec{e}_x$ in their plane with $\hat{\vec{\omega}}_{\rm R}\cdot\vec{e}_x>0$, and $\vec{e}_y=\vec{e}_z\times\vec{e}_x$ -- the mapping reads
\begin{align}
    \hat{\vec{\omega}}_+&=\frac{t_\mathrm{sc}\bar{\omega}\vec{e}_z-t_\mathrm{sf}\delta\omega\vec{e}_x}{t_c}\,,\qquad
    \hat{\vec{\omega}}_-=\frac{t_\mathrm{sc}\delta\omega\vec{e}_z-t_\mathrm{sf}\bar{\omega}\vec{e}_x}{t_c}\,,
    \label{eq:smdictpm}\\
    \hat{\vec{\omega}}_{\rm T}&=\omega_{{\rm T},x}^{\rm LZ}\vec{e}_x+\frac{t_\mathrm{sc}\omega_{{\rm T},y}^{\rm LZ}+t_\mathrm{sf}\nu_{\rm T}^{\rm LZ}}{t_c}\vec{e}_y+\omega_{{\rm T},z}^{\rm LZ}\vec{e}_z\,,\qquad
    \hat{\nu}_{\rm T}=\frac{t_\mathrm{sc}\nu_{\rm T}^{\rm LZ}-t_\mathrm{sf}\omega_{{\rm T},y}^{\rm LZ}}{t_c}\,,
    \label{eq:smdictMT}
\end{align}
where $\omega_{{\rm T},i}^{\rm LZ}$ are the components of $\vec{\omega}_{\rm T}^{\rm LZ}=\mu_Bg_\mathrm{T}^{\rm LZ}\vec{B}/\hbar$ in the local-Zeeman representation and $\nu_{\rm T}^{\rm LZ}=\boldsymbol{\mu}_{\rm T}^{\rm LZ}\cdot\vec{B}/\hbar$. Note the structure of Eq.~\eqref{eq:smdictMT}: the components transverse to $\vec{e}_y$ carry over unchanged, while the pair $(\omega_{{\rm T},y}^{\rm LZ},\nu_{\rm T}^{\rm LZ})$ is rotated by the half-angle $\widehat{\Theta}_{\rm LR}/2$ -- the same vector--scalar mixing of the magnetotunneling amplitudes that makes $g_\mathrm{T}^{\rm LZ}$ and $\boldsymbol{\mu}_{\rm T}^{\rm LZ}$ cumbersome, here reduced to a single angle. Substituting Eqs.~\eqref{eq:smdictpm} and~\eqref{eq:smdictMT} into Eqs.~\eqref{eq:smws} and~\eqref{eq:smgsvec}, the Larmor and spin-photon coupling vectors of the lower branch become
\begin{align}
    \hat{\vec{\omega}}_s&=\frac{t_\mathrm{sf}}{t_c}\left(\cos\Theta\bar{\omega}-\delta\omega\right)\vec{e}_x+\frac{t_\mathrm{sc}}{t_c}\left(\bar{\omega}-\cos\Theta\delta\omega\right)\vec{e}_z+\sin\Theta\hat{\vec{\omega}}_{\rm T},\\
    \hat{\vec{g}}_s&=-\frac{\hbar g_{\rm c}}{2}\left[\frac{1}{\Omega}+\frac{\Omega}{\Omega^2-(\hbar\omega_{\rm r})^2}\right]\sin\Theta\left[\sin\Theta\frac{t_\mathrm{sc}\delta\omega\vec{e}_z-t_\mathrm{sf}\bar{\omega}\vec{e}_x}{t_c}+\cos\Theta\hat{\vec{\omega}}_{\rm T}\right]\,,
    \label{eq:smLZvecs}\\
    \hat{\vec{\omega}}_\text{ad}&=2(\hbar g_{\rm c})^2\sin^2\Theta\left[\frac{1}{(\Omega-\hbar\omega_{\rm r})^2}+\frac{1}{(\Omega+\hbar\omega_{\rm r})^2}\right]\left(\cos\Theta\frac{t_\mathrm{sc}\delta\omega\vec{e}_z-t_\mathrm{sf}\bar{\omega}\vec{e}_x}{t_c}-\sin\Theta\hat{\vec{\omega}}_{\rm T}\right)\nonumber\\
    &\quad+\frac{8(\hbar g_{\rm c})^2\sin\Theta\cos\Theta}{\Omega^2-(\hbar\omega_{\rm r})^2}\left(\sin\Theta\frac{t_\mathrm{sc}\delta\omega\vec{e}_z-t_\mathrm{sf}\bar{\omega}\vec{e}_x}{t_c}+\cos\Theta\hat{\vec{\omega}}_{\rm T}\right)\,,
    \label{eq:smwadLZ}
\end{align}
where $\hat{\vec{\omega}}_s$ is written up to the $O(g_{\rm c}^2)$ corrections of Eq.~\eqref{eq:smws}. They map identically; the projections~\eqref{eq:smproj} apply unchanged. At zero detuning, Eqs.~\eqref{eq:smLZvecs} reproduce Eq.~\eqref{eq:smso0ang}; at the symmetric operating point $\varepsilon=\delta\omega=0$ and neglecting magnetotunneling, $\hat{\vec{\omega}}_s=(t_\mathrm{sc}/t_c)\bar{\omega}\vec{e}_z$ and $\hat{\vec{g}}_s=(\hbar g_{\rm c}\Lambda_0/2)(t_\mathrm{sf}/t_c)\bar{\omega}\vec{e}_x$ are exactly orthogonal: the coupling is purely transverse, $g_\parallel=0$ and $g_s=(\hbar g_{\rm c}\Lambda_0/2)\bar{\omega}t_\mathrm{sf}/t_c$, in agreement with the quasi-symmetric limit of section~\ref{sec:soframe} and, in the adiabatic limit, with Eqs.~(3) and~(4) of the main text. The local-Zeeman frame thus makes the flopping-mode structure of the problem explicit~\cite{yu2022strong}: the transverse spin-photon coupling is the spin-flip tunneling $t_\mathrm{sf}$ itself, dressed by the charge-photon coupling $g_{\rm c}$ and filtered by the charge dynamics through $\Lambda_0$.

As mentioned in section \ref{sec:insertion}, ${\omega}_s=(t_\mathrm{sc}/t_c)\bar{\omega}$ vanishes if $t_\mathrm{sc}\to0$ ($t_\mathrm{sf}\to t_c$): increasing $t_\mathrm{sf}/t_c$ generally enhances spin-photon coupling up to the limit where the magnetic field needed to achieve the target spin frequency $\omega_s$ becomes unreasonable (given, in particular, the presence of superconducting materials in the resonator). Targeting $t_\mathrm{sf}=0.8$ to $0.9t_c$ therefore appears as a safe optimum for most applications.

\bibliography{biblio.bib}

\end{document}